\ifx\pdfsuppresswarningpagegroup\undefined\else\pdfsuppresswarningpagegroup=1\fi\relax
\RequirePackage{silence}
\PassOptionsToPackage{unicode, pdfprintscaling=None, colorlinks}{hyperref}
\documentclass[ 10pt, twocolumn, english, aps, pra, longbibliography, superscriptaddress, amsmath, amssymb, floatfix]{revtex4-2}

\usepackage{dcolumn}
\usepackage{bm}
\usepackage{physics}
\usepackage{subcaption}
\DeclareCaptionJustification{justified}{\leftskip=0pt\rightskip=0pt\parfillskip=0ptplus1fil\relax}

\usepackage{quantikz}
\definecolor{riverlane_green}{RGB}{0, 150, 143}

\usepackage{soul}
\usepackage{amsmath,amssymb}
\usepackage{braket}
\usepackage{graphicx}
\usepackage{hyperref}
\usepackage{booktabs}
\usepackage[dvipsnames]{xcolor}
\usepackage{multirow} 
\usepackage{tikz}
\usetikzlibrary{positioning, arrows.meta, decorations.pathreplacing, calc, shapes.misc}

\definecolor{gorange}{HTML}{FCE5CD}
\definecolor{gobord}{HTML}{D6B656}
\definecolor{ggreen}{HTML}{D5E8D4}
\definecolor{ggbord}{HTML}{82B366}
\definecolor{gpur}{HTML}{E1D5E7}
\definecolor{gpbord}{HTML}{9673A6}
\definecolor{tcol}{HTML}{AA3377}
\definecolor{wg}{HTML}{888888}
\definecolor{mblue}{HTML}{4477AA}
\definecolor{mbbord}{HTML}{2A5A8A}
\definecolor{mcyan}{HTML}{66CCEE}
\definecolor{mcbord}{HTML}{449EBB}
\definecolor{mred}{HTML}{EE6677}
\definecolor{mrbord}{HTML}{CC4455}
\definecolor{mgray}{HTML}{F0F0F0}
\definecolor{gblue}{HTML}{BFDBFE}
\definecolor{gbbord}{HTML}{1D4ED8}

\makeatletter
\newcommand\showtitleinbib{{\escapechar=`\\ \immediate\write\@auxout{%
\csname citation{REVTEX42Control}\endcsname^^J%
\csname citation{apsrev42Control}\endcsname
}}}

\begin{document}

\preprint{APS/123-QED}

\title{\textbf{Learning Low-Energy Subspace Overlaps in Many-Body Systems with Measurement-Based and Coherent Quantum Strategies} 
}%

\author{Shamminuj Aktar}
\email{saktar@lanl.gov} 
\affiliation{Computing and Artificial Intelligence Division (CAI-3), Los Alamos National Laboratory, Los Alamos, New Mexico 87545, USA}

\author{Rishabh Bhardwaj}
\affiliation{Computing and Artificial Intelligence Division (CAI-3), Los Alamos National Laboratory, Los Alamos, New Mexico 87545, USA}

\author{Tanmoy Bhattacharya}
\affiliation{Theoretical Division (T-2), Los Alamos National Laboratory, Los Alamos, New Mexico 87545, USA}

\author{Stephan Eidenbenz}
\affiliation{Computing and Artificial Intelligence Division (CAI-3), Los Alamos National Laboratory, Los Alamos, New Mexico 87545, USA}

\date{\today}
             
%

\begin{abstract}
    Predicting the overlap of quantum states with specified low-energy subspaces is a key diagnostic for quantum many-body dynamics, with direct applications in state preparation, subspace-based algorithms, and the study of thermalization.
    We study the supervised prediction of subspace overlaps $O_K$ between time-evolved states and $K$-dimensional low-energy eigenspaces of a 10-qubit Heisenberg spin chain following a local perturbation. We compare two quantum information extraction strategies: measurement-based learning, in which classical shadow features are processed by convolutional neural networks, and coherent quantum learning, in which quantum convolutional neural networks process the state directly. We further introduce physics-informed variants for both approaches, including Hamiltonian-aware shadows and QCNN gates aligned with the Heisenberg exchange structure.
    Across five dataset configurations spanning weak, moderate, and strong quench regimes, physics-informed QCNNs achieve stable performance, with mean test-set coefficients of determination $R^2 = 0.753$--$0.846$. Shadow-based methods show stronger regime dependence: they outperform QCNNs in the moderate-quench regime, reaching $R^2 = 0.886$, but underperform in weak and strong quenches at default shot budgets, where the best shadow results are $R^2 = 0.615$ and $0.672$, respectively.
    Hardware validation on Quantinuum and IBM noise models shows that arbitrary state preparation is the dominant limitation, requiring approximately 2{,}044 two-qubit gates and causing near-complete depolarization before inference. These results identify a regime-dependent tradeoff between measurement-based and coherent quantum learning, with shadow methods excelling when the target remains locally accessible and physics-informed QCNNs providing more robust performance across dynamical regimes.
\end{abstract}

\maketitle
\section{Introduction}
    Quantum machine learning (QML) has emerged as a promising framework for addressing challenging problems in quantum many-body physics, particularly in settings where near-term quantum hardware may provide useful computational resources~\cite{biamonte2017quantum, preskill2018quantum, bharti2022noisy}. Variational algorithms and quantum learning methods have been applied for estimating ground- and excited-state energies~\cite{cerezo2021variational, henk2023ground, bao2025learning, patel2024quantum}, learning quantum phases~\cite{cong2019quantum, aktar2025quantum}, and benchmarking against classical techniques such as tensor networks, quantum Monte Carlo, and neural-network quantum states~\cite{bowles2024better, alvarez2025benchmarking, huang2022quantum}. These approaches are particularly relevant for condensed matter physics and quantum field theory, where strongly correlated dynamics and computational complexity are central. However, most efforts have focused on \emph{static} properties such as energies, order parameters, and equilibrium phase diagrams, leaving dynamical observables comparatively less explored.

    A related yet distinct class of observables carries significant physical information about quantum systems.
    These observables consist of overlap and transition amplitudes: quantities defined by inner products between states evolved under different Hamiltonians or projected onto specified spectral subspaces. They capture key features such as coherence, decoherence, and dynamical response~\cite{jalabert2001decoherence, zunkovic2016dynamical, heyl2018dynamical, heyl2013dynamical}. These amplitudes arise across multiple domains: for example, in high-energy physics as scattering amplitudes related to measurable cross sections~\cite{brown1992qft, preskillqft} and in condensed matter studies as Loschmidt amplitudes~\cite{rylands2019loschmidt, wong2022loschmidt}, which characterize quantum quenches, dynamical phase transitions, and scrambling. The \emph{overlap with a specified low-energy subspace} quantifies how much of a time-evolved state remains within a low-energy manifold of interest. It generalizes ground-state fidelity and can be used to track spectral redistribution during the crossover from perturbative to thermalizing dynamics~\cite{deutsch1991quantum, srednicki1994chaos, rigol2008thermalization, dalessio2016from}.

    Despite their importance, such overlaps are challenging to compute classically, especially when the dimension of the low-energy subspace of interest is large. Perturbative methods break down as interactions strengthen, quantum Monte Carlo often suffers from the sign problem in real-time dynamics~\cite{troyer2005computational, werner_noneq_qmc}, and tensor-network methods are limited when entanglement growth is rapid~\cite{mello2025clifford}. A limited number of works have targeted related overlap observables on quantum hardware. H{\'e}mery~\emph{et~al.}~\cite{hemery2024measuring} measured Loschmidt amplitudes for the Fermi-Hubbard model on a trapped-ion processor using a hybrid time-series algorithm. Related proposals for computing real-time scattering amplitudes in strongly interacting quantum field theories~\cite{briceno2024toward, li2024scattering} rely on fault-tolerant quantum computing architectures. In classical machine learning, overlaps typically appear as \emph{derived} quantities from full-wavefunction simulations or learned time evolution~\cite{gutierrez2022realtime, medvidovic2023variational, cincio2018learning}, rather than as primary learning targets. This leaves open how different quantum information extraction strategies perform when the overlap itself is the supervised learning target.
    
   We address this question by treating the low-energy subspace overlap as a \emph{primary learning target} and comparing two complementary strategies. The first is \emph{measurement-based}: randomized measurements produce classical shadow representations of the quantum state, which are then processed by a classical neural network~\cite{aaronson2018shadow, huang2020predicting, elben2023randomized}. Prior shadow work has focused on estimating fidelities and low-weight observables. Here, the subspace overlap
   $
    O_K = \sum_{\psi_n\in {\cal S}_K} |\langle\psi_n|\psi(t)\rangle|^2,
    $
    where $|\psi(t)\rangle$ is the time-evolved state and $\{|\psi_n\rangle\}$ form a basis for the specified $K$-dimensional low-energy subspace ${\cal S}_K$, is learned through a regression model from shadow-derived features rather than directly estimated from measurements. The second strategy is \emph{coherent}: quantum convolutional neural networks (QCNNs)~\cite{cong2019quantum, mangini2021quantum} process the quantum state directly through alternating convolutional and pooling layers. These architectures exhibit polynomially scaling gradients in relevant settings~\cite{pesah2021absence}, avoiding the barren plateaus encountered in generic parametrized circuits~\cite{mcclean2018barren, cerezo2021cost}. While QCNNs have been demonstrated experimentally for phase recognition~\cite{herrmann2022realizing, aktar2025quantum}, whether they offer advantages over classical post-processing of quantum measurements remains an open question~\cite{bermejo2024quantum}.

   Both approaches can be enhanced through physics-informed design. The Hamiltonian variational ansatz~\cite{wiersema2020exploring} and equivariant quantum neural networks~\cite{nguyen2024theory, meyer2023exploiting, larocca2022group} show that encoding structure into circuit architectures can improve optimization and generalization. We leverage this idea by introducing a Heisenberg QCNN whose convolutional gates reflect the underlying exchange interactions of the Hamiltonian. In parallel, we implement a Hamiltonian-aware shadow protocol that pre-rotates states before measurement. This enables a direct comparison of how Hamiltonian structure benefits measurement-based and coherent strategies.  We emphasize that these ``Hamiltonian''-aware methods build in the structure of the unperturbed Hamiltonian whose low-energy subspace we are studying in the learning protocol: the training protocol is still blind to the perturbations that produce the overlap that we intend to learn.
    
    Concretely, we study QCNN and shadow-fed classical CNNs architectures for predicting subspace overlaps in a ten-qubit one-dimensional Heisenberg spin chain following a local quench. The system is initialized in the ground state of a Hamiltonian $H_0$ and then evolved under a perturbed Hamiltonian $H = H_0 + H_1$, where $H_1$ is a local perturbation. The resulting time-evolved state $\ket{\psi(t)} = e^{-iHt}\ket{\psi_0}$, where $\ket{\psi_0}$ is the ground state of $H_0$, redistributes its weight across the energy spectrum. We predict the fraction of the evolved state remaining within a $K$-dimensional low-energy subspace ${\cal S}_K$ of $H_0$. Our results show that shadow measurements and coherent quantum processing provide complementary advantages across dynamical regimes. Our contributions are as follows:
    
    \begin{itemize}
        \item We compare shadow-based CNNs and QCNNs for predicting low-energy subspace overlaps across five datasets spanning three quench regimes and two subspace configurations, providing a systematic comparison of measurement-based and coherent strategies for this task.
        
        \item We introduce physics-informed architectures for both families: a Heisenberg QCNN and Heisenberg QCNN+ with Hamiltonian-aligned gates, and a Hamiltonian-aware shadow protocol that applies an $H_0$-based pre-rotation before measurement.
        
        \item Encoding Hamiltonian structure into QCNNs significantly improves performance, raising $R^2$ from $0.647$ to $0.832$ in the weak-quench regime, with consistent gains across all datasets. In contrast, the Hamiltonian-aware shadow protocol does not consistently outperform Pauli shadows and has the lowest performance in the strong-quench regime ($R^2 = 0.241$ at 500 shots, reaching $0.489$ at 5{,}000 shots).
        
        \item Physics-informed QCNNs achieve stable accuracy across all datasets ($R^2 = 0.753$--$0.846$), while shadow methods perform well only in the moderate regime, reaching a peak $R^2 = 0.886$, but dropping to $0.615$ and $0.672$ in weak and strong quenches. At higher shot budgets, Clifford shadows improve in the strong-quench regime, reaching $R^2 = 0.847$ at 5{,}000 shots.
        
        \item Noisy-backend experiments using IBM superconducting and Quantinuum trapped-ion models~\cite{kim2023evidence, moses2023race} show that arbitrary state preparation dominates the error budget (${\sim}2{,}044$ two-qubit gates), compared with the QCNN inference circuit (${\sim}150$ two-qubit gates).
    \end{itemize}
    
The rest of the paper is organized as follows. Section~\ref{sec:problem} defines the Heisenberg quench model, the low-energy subspace overlap target, and the dataset construction. Section~\ref{sec:methods} presents the measurement-based shadow protocols and coherent QCNN architectures, including the physics-informed variants. Section~\ref{sec:training} describes the training and evaluation procedure. Section~\ref{sec:results} compares the learning performance across quench regimes and measurement budgets. Section~\ref{sec:hardware} presents hardware validation on Quantinuum and IBM backends and analyzes the state-preparation bottleneck. Section~\ref{sec:conclusion} summarizes the main findings and discusses future directions.

\begin{figure*}[t!]
        \centering
        \includegraphics[width=0.95\textwidth]{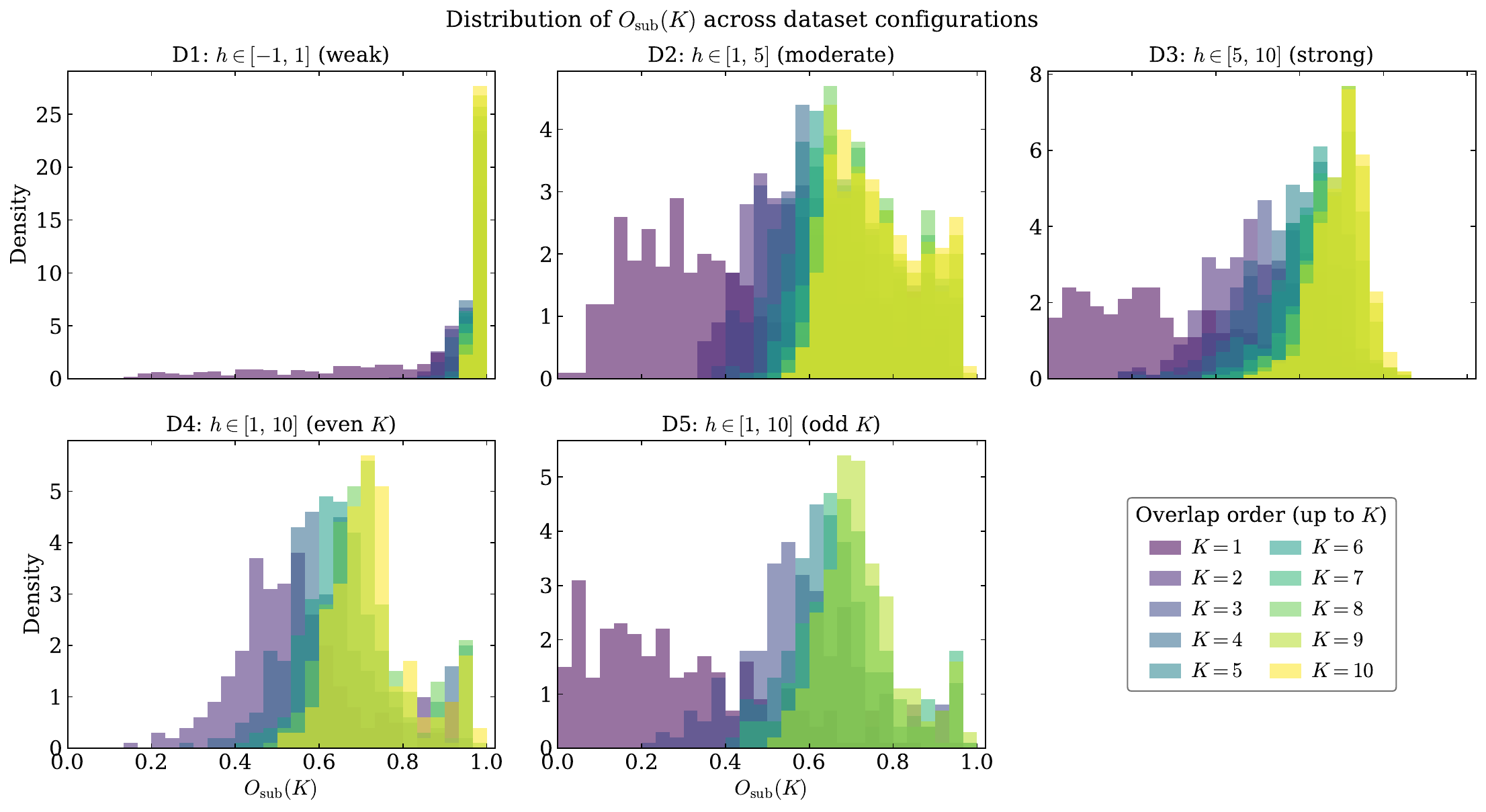}
       \caption{Distributions of $O_K$ across the five dataset configurations. Overlaps concentrate near unity in the weak-quench regime (D1) and spread toward lower values as the quench strength increases (D2, D3). Datasets D4 and D5 share the same quench range as D2--D3 but restrict the target overlap orders to even and odd values of $K$, respectively.}
        \label{fig:five-datasets}
    \end{figure*}
    \begin{table}[b!]
    \centering
        \caption{Dataset configurations. All share $N = 10$, $J = 1$, $h_z = -0.5$, $t \in [1, 20]$ with $\Delta t = 1.0$ (i.e., 20 equispaced steps), and $S = 300$ samples, where the symbols are defined in the text. Datasets D1--D3 probe distinct quench-strength regimes at all $K$; D4 and D5 span the full quench range with restricted $K$-sets to test generalization across overlap orders.}
        \label{tab:datasets}
        \begin{tabular}{@{}ccccc@{}}
            \toprule
           Dataset & $h$-range & $t$-range & $K$ values & Regime \\
            \midrule
            D1 & $[-1,\,1]$   & $[1,\,20]$ & $1,2,\dots,10$ & weak quench \\
            D2 & $[1,\,5]$    & $[1,\,20]$ & $1,2,\dots,10$ & moderate quench \\
            D3 & $[5,\,10]$   & $[1,\,20]$ & $1,2,\dots,10$ & strong quench \\
            D4 & $[1,\,10]$   & $[1,\,20]$ & $2,4,6,8,10$   & full range, even $K$ \\
            D5 & $[1,\,10]$   & $[1,\,20]$ & $1,3,5,7,9$    & full range, odd $K$ \\
            \bottomrule
        \end{tabular}
    \end{table}

\section{Problem Statement and Data Generation}
\label{sec:problem}

    \subsection{Heisenberg Model and Quench Protocol}
    \label{sec:model}
        We consider a supervised learning problem in which the goal is to learn \emph{low-energy subspace overlaps} of a time-evolved quantum many-body state following a local quench. Let $H_0$ denote the Hamiltonian of a one-dimensional spin-$\tfrac{1}{2}$ Heisenberg chain of length $N$ with open boundary conditions,
        \begin{equation}
            \label{eq:H0}
            H_0 = J \sum_{i=1}^{N-1}
            \bigl( X_i X_{i+1} + Y_i Y_{i+1} + Z_i Z_{i+1} \bigr) + h_z \sum_{i=1}^{N} Z_i ,
        \end{equation}
        where $X_i$, $Y_i$, and $Z_i$ are Pauli operators on site $i$, $J$ is the exchange coupling, and $h_z$ is a uniform longitudinal field. The isotropic Heisenberg exchange in Eq.~\eqref{eq:H0} is a canonical model of quantum magnetism~\cite{auerbach1994interacting} and a standard benchmark for quantum simulation and many-body algorithms~\cite{georgescu2014quantum}. Let $\{\ket{\psi_n}\}_{n=0}^{2^N - 1}$ denote the eigenstates of $H_0$, indexed by $n$ and ordered by increasing eigenvalue,
        \begin{equation}
            \label{eq:H0-eigenproblem}
            H_0 \ket{\psi_n} = E_n \ket{\psi_n} ,
        \end{equation}
        so that $\ket{\psi_0}$ is the ground state. A local quench is applied by adding a transverse field of strength $h$ at site $i_0$, yielding the post-quench Hamiltonian
        \begin{equation}
            \label{eq:H}
            H = H_0 + H_1, \qquad H_1 = h\, X_{i_0}.
        \end{equation}

        \begin{figure*}[t!]
            \centering
            \includegraphics[width=0.9\textwidth]{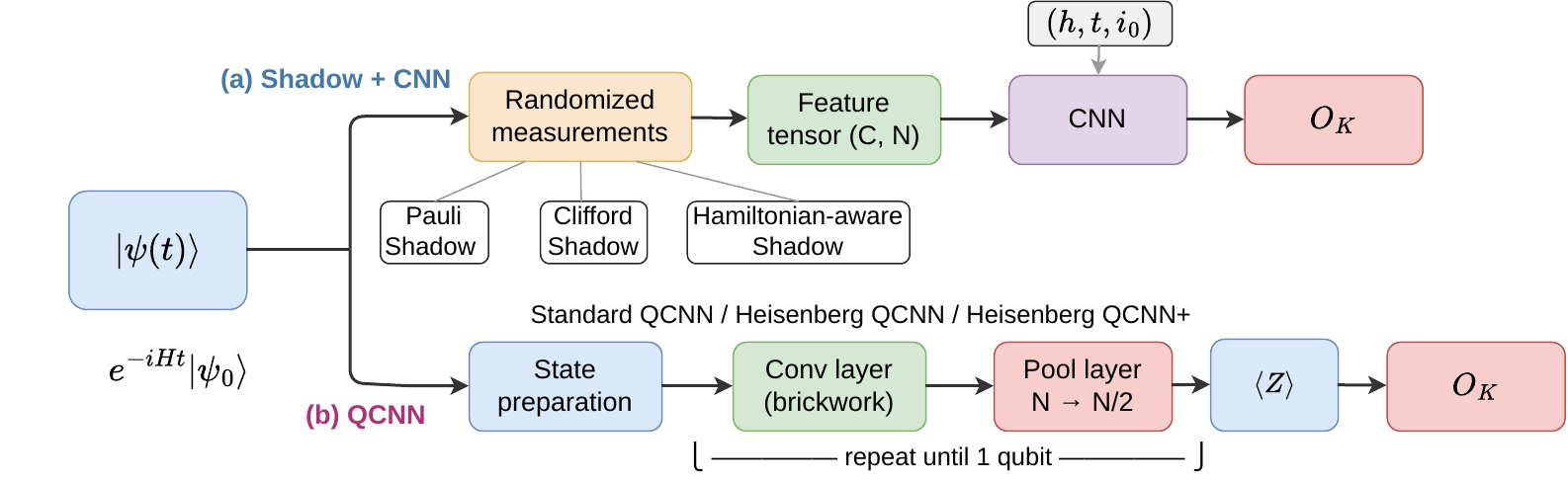}
            \caption{High-level comparison of the two quantum information extraction strategies. The input state $\ket{\psi(t)} = e^{-iHt}\ket{\psi_0}$ is the time-evolved ground state of $H_0$ under the post-quench Hamiltonian $H$. (a)~Measurement-based pipeline: one of three randomized measurement protocols (Pauli, Clifford, or Pauli on Hamiltonian-aware pre-rotation $H_0$-rot) produces a feature tensor of shape $(C, N)$, which is processed by a 1D CNN together with normalized scalar inputs $(h, t, i_0)$. (b)~Coherent pipeline: the statevector is loaded via arbitrary state preparation, processed through alternating brickwork convolutional and pooling layers that halve the active qubits at each stage, and the final $\langle Z \rangle$ is mapped to the overlap prediction via $(1 - \langle Z \rangle)/2$.}
            \label{fig:overview}
        \end{figure*}
        
        This local perturbation breaks the $U(1)$ spin-rotation symmetry of $H_0$ about the $z$-axis and, when $i_0$ is not at the chain center, spatial inversion symmetry. Starting from the ground state $\ket{\psi_0}$ of $H_0$, the system evolves unitarily under $H$:
        \begin{equation}
            \label{eq:psi-t}
            \ket{\psi(t)} = e^{-i H t}\, \ket{\psi_0} .
        \end{equation}
        Unlike a global quench, in which a system-wide parameter is changed~\cite{calabrese2007quantum, polkovnikov2011colloquium}, a local quench injects energy at a single site, producing entanglement that spreads through the chain~\cite{calabrese2007local, eisler2007evolution}. The strength $h$ controls how far the dynamics depart from the original low-energy subspace: small $h$ produces near-equilibrium perturbative behavior, while large $h$ drives stronger spectral redistribution associated with thermalizing dynamics and the Eigenstate Thermalization Hypothesis (ETH)~\cite{deutsch1991quantum, srednicki1994chaos, rigol2008thermalization}. For data generation, all time evolutions are computed by exact diagonalization:
        \[
        \ket{\psi(t)} = \sum_k e^{-i \epsilon_k t}\, c_k \ket{\phi_k},
        \]
        where $H\ket{\phi_k} = \epsilon_k \ket{\phi_k}$ and $c_k = \braket{\phi_k | \psi_0}$.

    \subsection{Overlap Target}
    \label{sec:target}
        Let ${\cal S}_K$ denote the subspace spanned by $K$ low-energy eigenstates of $H_0$. The rank-$K$ low-energy projector onto this subspace is
        \begin{equation}
            \label{eq:projector}
            P_K = \sum_{\psi_n\in{\cal S}_K} \ket{\psi_n}\!\bra{\psi_n} .
        \end{equation}
        The \emph{subspace overlap} is
        \begin{equation}
            \label{eq:Osub}
            O_K = \bra{\psi(t)} P_K \ket{\psi(t)}
            = \sum_{\psi_n\in{\cal S}_K} \bigl|\braket{\psi_n | \psi(t)}\bigr|^2 .
        \end{equation}
        For ${\cal S}_K=\{\psi_0\}$, this reduces to the ground-state fidelity, whose sensitivity to quantum phase transitions and critical phenomena is well established~\cite{zanardi2006ground, gu2010fidelity}. The generalization to $K > 1$ tracks how much spectral weight remains in the low-energy window as the quench drives the system away from the initial ground state~\cite{dalessio2016from, gogolin2016equilibration}. The learning task is to predict $O_K$ from $\ket{\psi(t)}$ for $K = 1,\dots,K_{\max}$, either processed directly by a quantum circuit or accessed through classical shadow measurements.

        \begin{figure*}[t!]
            \centering
            \resizebox{0.99\textwidth}{!}{

\begin{tikzpicture}[font=\sffamily\Large,
  pbg/.style={draw=mbbord, fill=mblue!25, minimum width=18mm, minimum height=14mm,
    rounded corners=2pt, font=\sffamily\LARGE\bfseries, inner sep=2pt},
  zbg/.style={draw=mbbord, fill=mblue!25, minimum width=14mm, minimum height=14mm,
    rounded corners=2pt, font=\sffamily\LARGE\bfseries, inner sep=2pt},
  cfg/.style={draw=mcbord, fill=mcyan!30, minimum width=21mm, minimum height=30mm,
    rounded corners=2pt, font=\sffamily\Large\bfseries, align=center, inner sep=2pt},
  isg/.style={draw=ggbord, fill=ggreen, minimum width=18mm, minimum height=26mm,
    rounded corners=2pt, font=\sffamily\Large\bfseries, align=center, inner sep=2pt},
  meas/.style={draw=mrbord, fill=mred!15, minimum width=14mm, minimum height=14mm,
    rounded corners=2pt, inner sep=2pt},
  cout/.style={draw=gray!80, fill=mgray, minimum width=22mm, minimum height=14mm,
    rounded corners=2pt, font=\sffamily\Large, inner sep=2pt},
]

\def\dy{1.6}
\def\ystart{-1.6}

\newcommand{\drawmeter}[2]{%
  \draw[mrbord, thick] (#1-0.22, #2-0.06) arc (180:0:0.22cm);
  \draw[mrbord, thick, -stealth] (#1, #2-0.06) -- ++(60:0.40cm);
}

\begin{scope}[shift={(0,0)}]
  \node[font=\sffamily\huge\bfseries, text=tcol] at (4.0,0.4)
    {(a) Pauli Shadow};

  \foreach \i in {0,...,3} {
    \pgfmathsetmacro{\yy}{\ystart - \i*\dy}
    \draw[wg, thick] (1.6, \yy) -- (5.6, \yy);
    \node[font=\sffamily\LARGE, text=black, anchor=east] at (1.5, \yy) {$q_{\i}$};
  }
  \pgfmathsetmacro{\ylast}{\ystart - 3*\dy}
  \node[font=\sffamily\LARGE, text=wg] at (1.3, \ylast-0.7) {$\vdots$};

  \draw[decorate, decoration={brace, mirror, amplitude=5pt}]
    (0.7, \ystart+0.5) -- (0.7, \ylast-0.5)
    node[midway, left=7pt, font=\sffamily\LARGE] {$\ket{\psi}$};

  \node[pbg] at (2.8, \ystart)        {$U_Z$};
  \node[pbg] at (2.8, \ystart-\dy)    {$U_X$};
  \node[pbg] at (2.8, \ystart-2*\dy)  {$U_Y$};
  \node[pbg] at (2.8, \ystart-3*\dy)  {$U_X$};

  \foreach \i in {0,...,3} {
    \pgfmathsetmacro{\yy}{\ystart - \i*\dy}
    \node[meas] at (4.6, \yy) {};
    \drawmeter{4.6}{\yy}
  }

  \draw[gray, dashed, thick] (5.5, \ystart+0.5) -- (5.5, \ylast-0.5);

  \pgfmathsetmacro{\ymid}{\ystart - 1.5*\dy}
  \draw[gray!70, very thick] (6.1, \ystart+0.3) -- (6.1, \ylast-0.3);
  \foreach \i in {0,...,3} {
    \pgfmathsetmacro{\yy}{\ystart - \i*\dy}
    \draw[gray!70, thick] (5.5, \yy) -- (6.1, \yy);
  }
  \draw[->, gray!70, thick] (6.1, \ymid) -- (6.6, \ymid);

  \node[cout] at (7.4, \ymid) {$\langle \hat{P}_i \rangle$};
  \draw[->, gray, thick] (8.4, \ymid) -- (8.9, \ymid);
  \node[font=\sffamily\LARGE, text=black, anchor=west] at (9.0, \ymid)
    {$(3, N)$};
\end{scope}

\begin{scope}[shift={(11,0)}]
  \node[font=\sffamily\huge\bfseries, text=tcol] at (4.6,0.4)
    {(b) 2q-Clifford Shadow};

  \foreach \i in {0,...,3} {
    \pgfmathsetmacro{\yy}{\ystart - \i*\dy}
    \draw[wg, thick] (1.6, \yy) -- (5.8, \yy);
  }
  \pgfmathsetmacro{\ylast}{\ystart - 3*\dy}
  \node[font=\sffamily\LARGE, text=wg] at (1.3, \ylast-0.7) {$\vdots$};

  \pgfmathsetmacro{\ymidA}{\ystart - 0.5*\dy}
  \pgfmathsetmacro{\ymidB}{\ystart - 2.5*\dy}
  \node[cfg] at (3.0, \ymidA) {$C_{\!2q}^{(0,1)}$};
  \node[cfg] at (3.0, \ymidB) {$C_{\!2q}^{(2,3)}$};

  \foreach \i in {0,...,3} {
    \pgfmathsetmacro{\yy}{\ystart - \i*\dy}
    \node[meas] at (5.0, \yy) {};
    \drawmeter{5.0}{\yy}
  }

  \draw[gray, dashed, thick] (5.9, \ystart+0.5) -- (5.9, \ylast-0.5);

  \pgfmathsetmacro{\ymid}{\ystart - 1.5*\dy}
  \draw[gray!70, very thick] (6.5, \ystart+0.3) -- (6.5, \ylast-0.3);
  \foreach \i in {0,...,3} {
    \pgfmathsetmacro{\yy}{\ystart - \i*\dy}
    \draw[gray!70, thick] (5.9, \yy) -- (6.5, \yy);
  }

  \pgfmathsetmacro{\youta}{\ystart - 0.9*\dy}
  \pgfmathsetmacro{\youtb}{\ystart - 2.1*\dy}
  \draw[->, gray!70, thick] (6.5, \youta) -- (7.0, \youta);
  \draw[->, gray!70, thick] (6.5, \youtb) -- (7.0, \youtb);

  \node[cout, minimum width=24mm] at (7.9, \youta) {$\langle \hat{P}_i \rangle$};
  \node[cout, minimum width=24mm] at (7.9, \youtb) {$\langle \hat{P}_i\hat{P}_j \rangle$};
  \draw[->, gray, thick] (9.0, \ymid) -- (9.5, \ymid);
  \node[font=\sffamily\LARGE, text=black, anchor=west] at (9.6, \ymid)
    {$(6, N)$};
\end{scope}

\begin{scope}[shift={(22.5,0)}]
  \node[font=\sffamily\huge\bfseries, text=tcol] at (6.2,0.4)
    {(c) Hamiltonian Shadow};

  \foreach \i in {0,...,3} {
    \pgfmathsetmacro{\yy}{\ystart - \i*\dy}
    \draw[wg, thick] (1.6, \yy) -- (10.4, \yy);
  }
  \pgfmathsetmacro{\ylast}{\ystart - 3*\dy}
  \node[font=\sffamily\LARGE, text=wg] at (1.3, \ylast-0.7) {$\vdots$};

  \draw[ggbord, dashed, thick, rounded corners=4pt]
    (1.7, \ystart+0.9) rectangle (6.5, \ylast-0.9);
  \node[font=\sffamily\huge\bfseries, text=ggbord, anchor=south]
    at (4.1, \ystart+0.9) {$e^{-i\tau H_0}$};

  \pgfmathsetmacro{\ymidA}{\ystart - 0.5*\dy}
  \pgfmathsetmacro{\ymidB}{\ystart - 1.5*\dy}
  \pgfmathsetmacro{\ymidC}{\ystart - 2.5*\dy}
  \node[isg] at (2.7, \ymidA) {$XX$\\$YY$\\$ZZ$};
  \node[isg] at (2.7, \ymidC) {$XX$\\$YY$\\$ZZ$};
  \node[isg] at (4.1, \ymidB) {$XX$\\$YY$\\$ZZ$};

  \foreach \i in {0,...,3} {
    \pgfmathsetmacro{\yy}{\ystart - \i*\dy}
    \node[zbg] at (5.7, \yy) {$Z$};
  }

  \node[pbg] at (7.9, \ystart)        {$U_Y$};
  \node[pbg] at (7.9, \ystart-\dy)    {$U_Z$};
  \node[pbg] at (7.9, \ystart-2*\dy)  {$U_X$};
  \node[pbg] at (7.9, \ystart-3*\dy)  {$U_Z$};

  \foreach \i in {0,...,3} {
    \pgfmathsetmacro{\yy}{\ystart - \i*\dy}
    \node[meas] at (9.6, \yy) {};
    \drawmeter{9.6}{\yy}
  }

  \draw[gray, dashed, thick] (10.5, \ystart+0.5) -- (10.5, \ylast-0.5);

  \pgfmathsetmacro{\ymid}{\ystart - 1.5*\dy}
  \draw[gray!70, very thick] (11.1, \ystart+0.3) -- (11.1, \ylast-0.3);
  \foreach \i in {0,...,3} {
    \pgfmathsetmacro{\yy}{\ystart - \i*\dy}
    \draw[gray!70, thick] (10.5, \yy) -- (11.1, \yy);
  }
  \draw[->, gray!70, thick] (11.1, \ymid) -- (11.6, \ymid);

  \node[cout] at (12.4, \ymid) {$\langle \hat{P}_i \rangle$};
  \draw[->, gray, thick] (13.4, \ymid) -- (13.9, \ymid);
  \node[font=\sffamily\LARGE, text=black, anchor=west] at (14.0, \ymid)
    {$(3, N)$};
\end{scope}

\end{tikzpicture}}
            \caption{Shadow measurement protocols. (a)~Pauli shadow: random single-qubit Pauli rotations yield $3N$ local expectation features via median-of-means (8~groups, 500~shots). (b)~Two-qubit Clifford shadow: random 2Q Clifford circuits in alternating brickwork pairing yield $6N$ features including nearest-neighbor correlators (8~groups, 1{,}000~shots). (c)~Hamiltonian-aware shadow: $e^{-i\tau H_0}$ pre-rotation consisting of nearest-neighbor Ising exchange gates ($XX$, $YY$, $ZZ$) and single-qubit $Z$ rotations representing the longitudinal field $h_z \sum_i Z_i$, followed by random Pauli measurements (8~groups, 500~shots).} 
            \label{fig:shadow-protocols}
        \end{figure*}
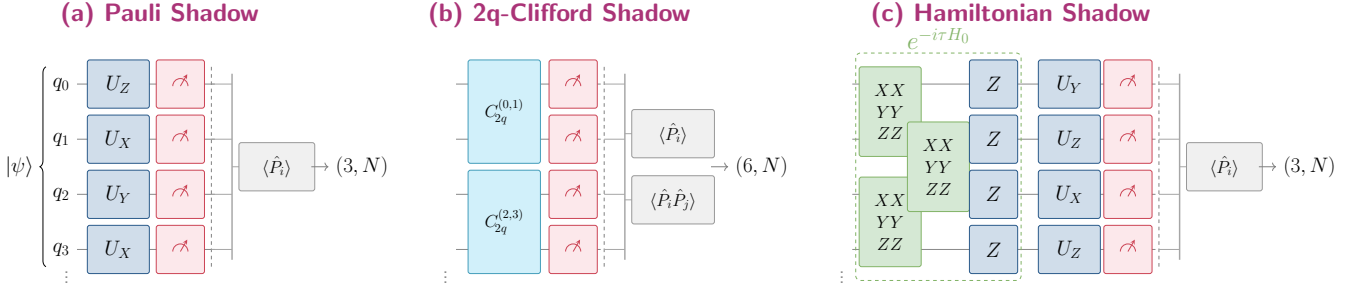
        
    \subsection{Dataset Generation}
    \label{sec:datasets}
        All datasets share the system parameters $N = 10$, $J = 1$, $h_z = -0.5$, and $K_{\max} = 10$. The target $O_K$ is cumulative, measuring the total weight of $\ket{\psi(t)}$ in the lowest $K$ eigenstates of $H_0$ included in ${\cal S}_K$, as defined in Eq.~\eqref{eq:Osub}. Five dataset configurations probe different aspects of the learning problem, as summarized in Table~\ref{tab:datasets}. Datasets $D_1$--$D_3$ separate weak ($D_1$), moderate ($D_2$), and strong ($D_3$) quench regimes, enabling analysis of how each learning paradigm responds to increasing departure from the equilibrium subspace. Datasets $D_4$ and $D_5$ instead test generalization across overlap orders on a broad quench range: $D_4$ targets even $K$ and $D_5$ targets odd $K$, covering $h \in [1, 10]$.
        
        For each dataset, a discrete grid of quench parameters is constructed from four interior quench sites drawn uniformly without replacement, excluding the boundary spins, $15$ equispaced quench strengths $h \in [h_{\min}, h_{\max}]$, and $20$ equispaced evolution times $t \in [t_{\min}, t_{\max}]$, yielding $4 \times 15 \times 20 = 1{,}200$ parameter combinations. The site selection is seeded independently per dataset. From this pool, $S = 300$ samples are drawn uniformly at random. This sample size balances coverage of the parameter space with the computational cost of exact diagonalization and repeated training across models and overlap orders. Each dataset is split $80/20$, with $240$ training samples and $60$ test samples, and split indices are fixed across all models and runs. Figure~\ref{fig:five-datasets} shows the resulting overlap distributions. In $D_1$, overlaps concentrate near unity for $K \geq 2$, with meaningful spread only at $K = 1$, reflecting that weak quenches barely displace spectral weight from the low-energy subspace. In $D_2$, distributions broaden across all $K$ as moderate quenches redistribute spectral weight, producing the largest variance across the three regimes. In $D_3$, overlaps shift toward lower values as strong quenches produce stronger spectral redistribution. Datasets $D_4$ and $D_5$ cover the combined quench range of $D_2$ and $D_3$, and their distributions reflect the corresponding spread.
    

\section{Quantum Information Extraction Methods}\label{sec:methods}
    We investigate two complementary families of methods for learning subspace overlaps. Measurement-based approaches extract classical features from randomized measurements and process them with a neural network, while coherent approaches operate directly on the quantum state using a parametrized quantum circuit. Both strategies access the same underlying quantum state but differ in how quantum information is processed.
    Measurement-based methods rely on statistical estimation from many shallow circuits, while coherent methods perform the computation in a single coherent execution. Fig.~\ref{fig:overview} provides a high-level comparison of the two pipelines, and Table~\ref{tab:model-summary} summarizes all seven architectures.

    \subsection{Measurement-based Methods: Classical Shadow Tomography}\label{sec:shadows}
        Classical shadow tomography~\cite{aaronson2018shadow, huang2020predicting} provides a framework for predicting many properties of an unknown quantum state from randomized measurements. A collection of random measurement outcomes, combined with the inverse of the measurement channel, yields an efficient classical representation called a classical shadow. From this representation, expectation values of many observables can be estimated simultaneously with provable sample-complexity guarantees~\cite{huang2020predicting}. All three shadow models in this study share a common pipeline. A measurement scheme produces a feature tensor from $\ket{\psi(t)}$, which is combined with normalized scalar inputs $(h, t, i_0)$ and then processed by a 1D CNN. The shadow features are independent of the overlap order $K$. They are computed once per state and reused across all $K$ values, which reduces computational cost.
        
        The Pauli Shadow measures each qubit in a uniformly random Pauli basis ($X$, $Y$, or $Z$), yielding local expectations $\langle X_i \rangle$, $\langle Y_i \rangle$, $\langle Z_i \rangle$ as a $(3, N)$ feature tensor~\cite{huang2020predicting}. This single-qubit protocol is the simplest and most noise-resilient shadow scheme, since each measurement circuit requires at most one single-qubit rotation.
        
        The Clifford Shadow applies random two-qubit Clifford unitaries in an alternating brickwork pattern, randomly choosing even or odd pairing per shot~\cite{hu2023classical}. This extracts both single-site expectations and nearest-neighbor correlators $\langle P_i P_{i+1} \rangle$ ($P \in \{X,Y,Z\}$) into a $(6, N)$ feature tensor. The two-qubit protocol captures entanglement structure, including information about two-qubit reduced states, that is inaccessible to Pauli shadows. This comes at the cost of increasing the measurement budget relative to Pauli shadows and requiring entangling gates in the measurement circuit.
        
        The Hamiltonian Shadow pre-rotates each state by $e^{-i\tau H_0}$ (where we choose $\tau = 1.0$) before applying standard Pauli measurements~\cite{hu2022hamiltonian}, where $H_0$ is the unperturbed Heisenberg Hamiltonian~[Eq.~\eqref{eq:H0}]. The pre-rotation is intended to align the measurement basis with the eigenstructure of $H_0$, making subsequent Pauli measurements more informative about low-energy spectral weight.
        
        All shadow features are estimated using a median-of-means estimator with 8~groups, which controls failure probability without requiring Gaussian concentration~\cite{huang2020predicting}. The measurement budgets are 500~shots for the Pauli and Hamiltonian shadows, and 1{,}000~shots for the Clifford shadow. The higher budget for Clifford measurements compensates for the additional statistical noise introduced by two-qubit random circuits. Fig.~\ref{fig:shadow-protocols} illustrates the three measurement protocols.
        
        \begin{figure*}[t!]
          \centering
          \resizebox{0.8\textwidth}{!}{\input{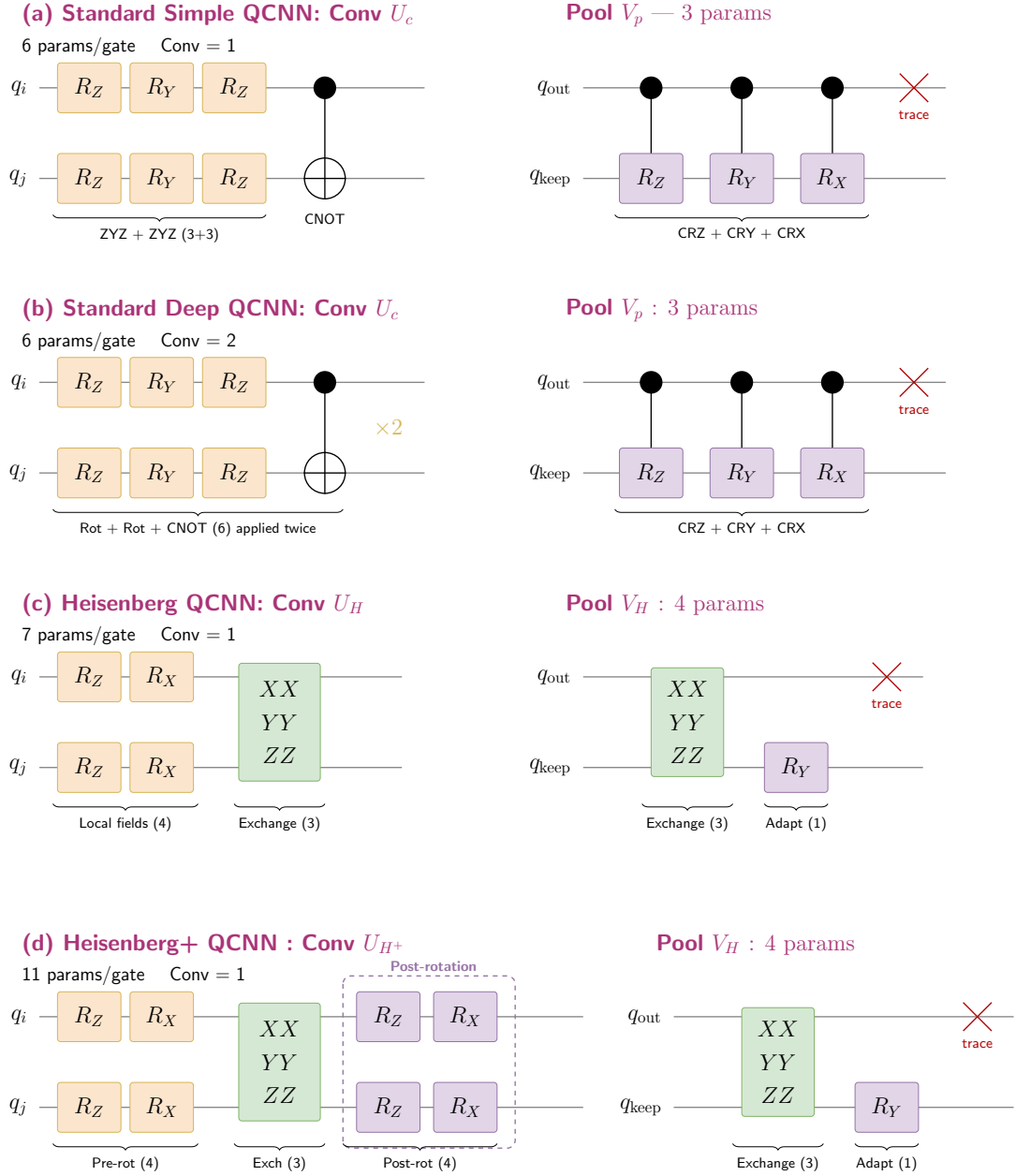}}
          \caption{Convolutional and pooling gate structures for the four QCNN variants. (a)~Standard Simple: $\mathrm{Rot}+\mathrm{CNOT}$ (6~params) with CRZ+CRY+CRX pooling (3~params). (b)~Standard Deep: same gate applied twice per block. (c)~Heisenberg QCNN: local rotations ($R_Z$, $R_X$) followed by Ising exchange ($XX$, $YY$, $ZZ$) mirroring $H_0$ (7~params), with exchange-based pooling and $R_Y$ (4~params). (d)~Heisenberg QCNN+: adds post-rotation for enhanced expressibility (11~params); same pooling as~(c).}
         \label{fig:conv-pool-gates}
        \end{figure*}

        The shared CNN architecture consists of three convolutional layers (32, 64, 128 filters; kernel size~3; batch normalization; ReLU activation), adaptive average pooling, dropout ($p = 0.2$), and a three-layer fully connected head $(128+3) \to 64 \to 32 \to 1$ with a sigmoid output activation. The $+3$ accounts for the three normalized scalar inputs $(h, t, i_0)$ concatenated before the fully connected head. Kernels of size~3 are suited to local patterns in the feature tensor, while the layer hierarchy builds longer-range representations. The Pauli and Hamiltonian Shadow models have 42{,}241 trainable parameters; the Clifford model has 42{,}529 due to its larger input channel count.
        \begin{table*}
            \centering
           \caption{Summary of the seven models ($N = 10$), showing default shot budgets, input size, and total parameter counts. Shadow features are precomputed once and shared across all $K$ values.}
            \label{tab:model-summary}
            \begin{tabular}{@{}llcrll@{}}
                \toprule
                Model & Family & Input & Params & Shots & Key design choice \\
                \midrule
                Pauli Shadow CNN       & Shadow+CNN    & $(3,N)$ & 42\,241 & 500   & Local Pauli expectations \\
                Clifford Shadow CNN    & Shadow+CNN    & $(6,N)$ & 42\,529 & 1\,000 & + NN correlators \\
                Hamiltonian Shadow CNN & Shadow+CNN    & $(3,N)$ & 42\,241 & 500   & $e^{-i\tau H_0}$ pre-rotation \\
                \midrule
                QCNN Simple            & QCNN    & $\ket{\psi}$ & 123  & exact & Generic gates, 1 conv/block \\
                QCNN Deep              & QCNN    & $\ket{\psi}$ & 219  & exact & Generic gates, 2 conv/block \\
                Heisenberg QCNN        & QCNN    & $\ket{\psi}$ & 148  & exact & Ising exchange gates \\
                Heisenberg QCNN+       & QCNN    & $\ket{\psi}$ & 212  & exact & Ising exchange gates + post-rotation \\
                \bottomrule
            \end{tabular}
        \end{table*}
        
    \subsection{Coherent Methods: Quantum Convolutional Neural Networks (QCNNs)}\label{sec:qcnns}
        QCNNs~\cite{cong2019quantum} process quantum states through alternating convolutional and pooling layers that progressively halve the number of active qubits, in analogy with the pooling operations of classical CNNs~\cite{lecun2015deep}. The final qubit's expectation value $\langle Z \rangle$ is mapped to the overlap prediction via $(1 - \langle Z \rangle)/2$. QCNNs exhibit favorable trainability properties, with polynomially scaling gradients in certain settings~\cite{pesah2021absence}, in contrast to generic parametrized circuits that can suffer from exponentially vanishing gradients~\cite{mcclean2018barren, cerezo2021cost,nagano2023quantum}.
    
        In the standard QCNN variants, each convolutional gate applies $\mathrm{Rot}(\phi,\theta,\omega)$ on each qubit, followed by a CNOT, giving 6 parameters per gate. Pooling uses controlled rotations (CRZ, CRY, CRX) from the traced-out qubit to the retained qubit (3 parameters). The QCNN Simple model applies one convolutional pass (even pairs followed by odd pairs) per block before pooling, resulting in 123 parameters. The QCNN Deep model applies two convolutional passes per block, increasing the parameter count to 219.
        
        The Heisenberg QCNN is the central architectural contribution of this work. Each convolutional gate consists of local rotations $R_Z$ and $R_X$ on each qubit (4~parameters), followed by a Heisenberg exchange block: $\mathrm{IsingXX}(\alpha)$, $\mathrm{IsingYY}(\beta)$, $\mathrm{IsingZZ}(\gamma)$ (3~parameters), directly reflecting the nearest-neighbor exchange terms in $H_0$ [Eq.~\eqref{eq:H0}]. Each gate implements $e^{-i\theta P_i P_{i+1}/2}$ for $P \in \{X, Y, Z\}$; explicit matrix forms are given in the Supplementary (Eqs.~\ref{eq:ising_unitary}--\ref{eq:ising_matrix}). This design follows the Hamiltonian Variational Ansatz~\cite{wiersema2020exploring}, where the Hamiltonian structure provides an inductive bias while all gate parameters remain trainable. The approach is further motivated by the theory of equivariant quantum neural networks~\cite{nguyen2024theory, meyer2023exploiting, larocca2022group}, which shows that encoding symmetries and problem structure into circuit gates can improve generalization. Concretely, the Ising exchange gates generate the same interaction algebra as the Heisenberg coupling, encouraging the circuit to operate within physically relevant sectors of $H_0$. Pooling gates use the same Ising exchange structure plus an $R_Y$ rotation on the kept qubit (4~parameters total). The Heisenberg QCNN has 148~parameters ($16 \times 7 + 9 \times 4$).
        
        The Heisenberg QCNN+ appends a learnable post-rotation ($R_Z$, $R_X$ per qubit; 4~additional parameters) after each exchange block, bringing each convolutional gate to 11~parameters and the total to 212. The post-rotation enhances expressibility by allowing the circuit to rotate into a basis better suited for subsequent layers, which is particularly relevant when the quench drives states far from the $H_0$ eigenbasis. Fig.~\ref{fig:conv-pool-gates} shows the gate structures for all four QCNN variants.

\section{Training Protocol}\label{sec:training}
    Models are trained using the Adam optimizer~\cite{kingma2017adammethodstochasticoptimization}. For shadow CNNs, we use a learning rate of $10^{-3}$ with $L_2$ weight decay $10^{-4}$ and mean squared error (MSE) loss. QCNNs use a learning rate of $10^{-2}$ and Huber loss ($\delta = 0.05$). Both model families employ early stopping and learning rate scheduling. Full hyperparameters are listed in Supplementary Table~\ref{tab:resources-hyperparams}.
    
    QCNN gradients are computed using adjoint differentiation on the \texttt{lightning.qubit} simulator~\cite{bergholm2018pennylane}. Circuit parameters are initialized from $\mathcal{N}(0,\,0.01^2)$ to mitigate barren plateaus~\cite{grant2019initialization}. In all simulations, both model families start from the same ideal noiseless quantum states, ensuring a controlled comparison at the level of the input state. The difference lies in how quantum information is accessed: QCNNs operate directly on the statevector $\ket{\psi(t)}$, whereas shadow CNNs use shot-sampled features obtained from randomized measurements of $\ket{\psi(t)}$.
    
    Shadow features are precomputed once per dataset and reused across all $K$ values. 
    Each model, dataset, and $K$ combination is trained over three independent runs. Performance is evaluated using the coefficient of determination $R^2$ on the held-out test set, reported as the mean across runs. The $R^2$ metric is defined as
    \begin{equation}
        \label{eq:r2}
        R^2 = 1 - \frac{\sum_i (y_i - \hat{y}_i)^2}{\sum_i (y_i - \bar{y})^2},
    \end{equation}
    where $y_i$ are the true overlap values, $\hat{y}_i$ are the model predictions, and $\bar{y}$ is the mean of the true values over the test set. This metric quantifies the fraction of variance explained by the model, with $R^2 = 1$ indicating perfect prediction and $R^2 = 0$ corresponding to a constant predictor. Additional details, including full hyperparameters, gradient computation, and evaluation metrics, are provided in the Supplementary Information (Training Details).

    \begin{figure*}[t!]
        \centering
        \includegraphics[width=0.95\textwidth]{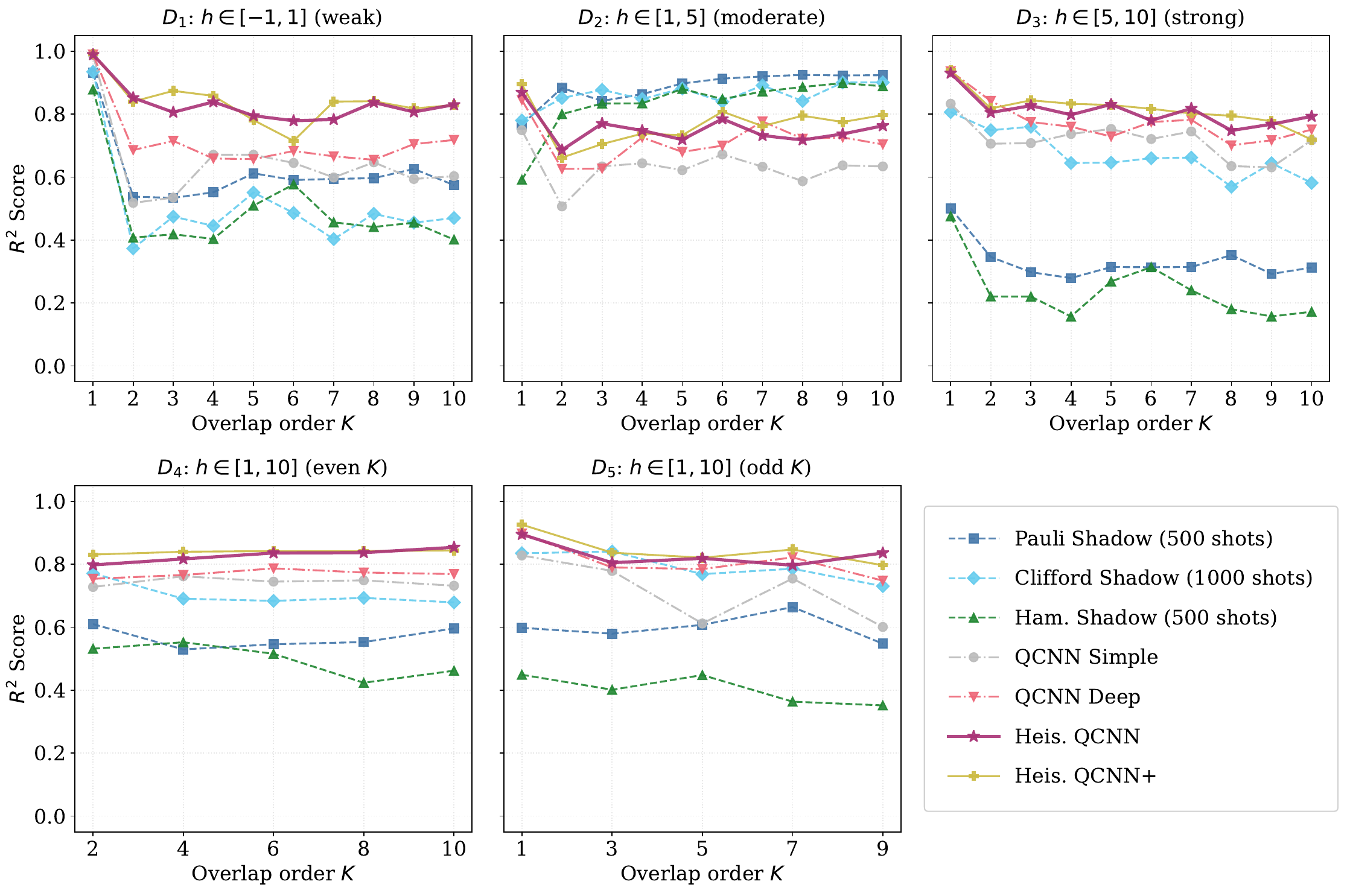}
        \caption{Test-set $R^2$ as a function of cumulative overlap order $K$ for all seven models across the five dataset configurations at the measurement budgets listed in Table~\ref{tab:model-summary}. Each panel shows one dataset: D1 (weak quench), D2 (moderate quench), D3 (strong quench), D4 (full range, even $K$), and D5 (full range, odd $K$). Values show mean $R^2$ averaged over three independent runs. Both measurement-based and coherent strategies successfully learn subspace overlaps, but relative performance depends strongly on the quench regime. Supplementary Table~\ref{tab:sup-results-summary} gives the corresponding $R^2$ values averaged over all $K$.}
        \label{fig:r2-vs-k}
    \end{figure*}

\section{Experimental Results}\label{sec:results}
    \subsection{Learning Subspace Overlaps}\label{sec:both-learn}
        Fig.~\ref{fig:r2-vs-k} shows the test-set $R^2$ (Eq.~\ref{eq:r2}) as a function of cumulative overlap order $K$ for all seven models across the five dataset configurations. All models achieve positive $R^2$ across all datasets and overlap orders, demonstrating that both measurement-based and coherent strategies can learn subspace overlaps. A full per-dataset summary is provided in Supplementary Table~\ref{tab:sup-results-summary}. The highest single-dataset score is achieved by the Pauli Shadow CNN, which reaches $R^2 = 0.886$ on $D_2$ at 500~shots. All QCNN variants show robust learning across all datasets, with Heisenberg QCNN+ reaching $R^2 = 0.839$ on $D_1$. Datasets $D_4$ and $D_5$ probe generalization across the full quench range $h \in [1,10]$ using even and odd $K$-sets, respectively. The results are consistent with the regime-specific trends: Heisenberg QCNN+ achieves the best performance on both datasets ($D_4$: $R^2 = 0.840$, $D_5$: $0.846$), Clifford Shadow is the strongest shadow model ($D_4$: $0.704$, $D_5$: $0.792$), and Hamiltonian Shadow has the lowest performance ($D_4$: $0.497$, $D_5$: $0.403$). These results indicate that the optimal learning strategy depends on the quench regime, as explored in the following subsections.

        \begin{figure*}[t!]
            \centering
            \includegraphics[width=0.95\textwidth]{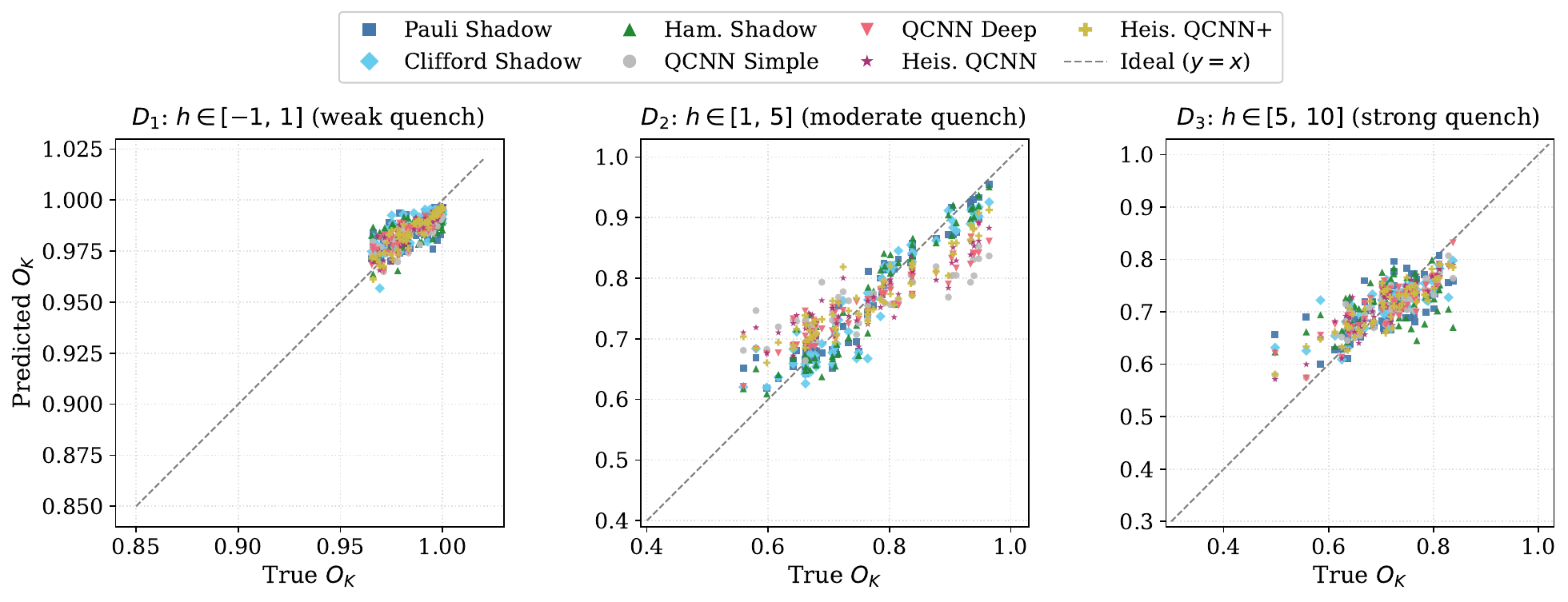}
            \caption{Predicted versus true overlap $O_K$ at $K = 10$ for all seven models across $D_1$, $D_2$, and $D_3$. Each point represents one test sample. Points along the diagonal indicate accurate predictions. In the weak-quench regime ($D_1$), Heisenberg QCNN and Heisenberg QCNN$^+$ cluster tightly along the diagonal while shadow model points show greater scatter. In the moderate-quench regime ($D_2$), all three shadow models align closely with the diagonal, outperforming the QCNNs. In the strong-quench regime ($D_3$), Heisenberg QCNN produces the tightest alignment while Pauli and Hamiltonian shadow predictions are broadly scattered.}
            \label{fig:scatter}
        \end{figure*}
        
    \subsection{Performance across quench regimes}\label{sec:regimes}
        Each quench regime presents qualitatively different physical conditions, and the two strategies learn from them differently. In $D_1$, meaningful variation appears mainly at $K=1$, where the target depends on fine low-energy spectral structure that local measurements do not capture well. Fig.~\ref{fig:r2-vs-k} shows that both Heisenberg QCNN variants perform well, consistent with their exchange gates providing a useful inductive bias aligned with $H_0$. Heisenberg QCNN+ achieves $R^2 = 0.839$, while the best shadow model reaches only $0.615$.
        
      In $D_2$, spectral weight redistribution appears sufficiently reflected in local observables that single-site Pauli expectations retain predictive power. Shadow models learn this structure efficiently, with all three surpassing all four QCNN variants. Pauli Shadow CNN reaches $R^2 = 0.886$ at 500~shots against the best QCNN score of $0.767$.
        
       In $D_3$, stronger quench dynamics make local observables less informative. Single-qubit measurements cannot capture the multi-qubit correlations that carry predictive signal, so Pauli and Hamiltonian shadows fall to $R^2 = 0.332$ and $0.241$. Clifford Shadow, which includes nearest-neighbor correlator information through two-qubit randomized measurements~\cite{hu2023classical}, reaches $R^2 = 0.672$. All QCNN variants learn consistently, with Heisenberg QCNN+ at $R^2 = 0.817$.
        
       Overall, shadow methods underperform in $D_1$, where the target is not well captured by local features, and in $D_3$, where stronger quench dynamics require information beyond the default measurement budgets. In $D_2$, the relevant features remain more accessible to shadow measurements, leading to stronger performance. QCNNs are more robust across all three regimes because coherent processing with physics-aligned gates is less constrained by local measurement features. Fig.~\ref{fig:scatter} further illustrates this behavior through predicted-versus-true scatter plots at $K = 10$. 
         \begin{figure*}[t!]
            \centering
            \includegraphics[width=0.99\textwidth]{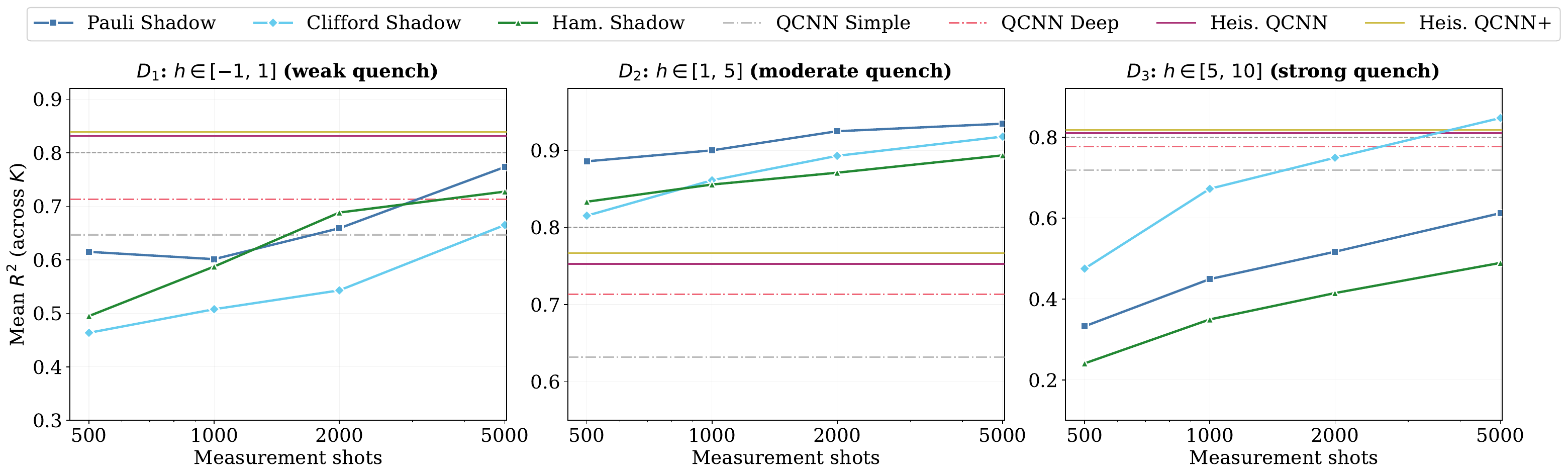}
            \caption{Mean $R^2$ averaged across all $K$ values as a function of measurement shot count for the three shadow models on $D_1$, $D_2$, and $D_3$. QCNN models appear as shot-independent horizontal lines and the gray dashed line marks $R^2 = 0.8$. On $D_2$, all shadow models surpass every QCNN already at 500 shots, with Pauli Shadow reaching $R^2 = 0.934$ at 5{,}000 shots. On $D_3$, Clifford Shadow is the only shadow variant to cross the QCNN threshold, reaching 0.847 at 5{,}000 shots against the Heisenberg QCNN$^+$ baseline of 0.817, while Pauli and Hamiltonian shadows reach only 0.612 and 0.489 at 5{,}000 shots. On $D_1$, no shadow model reaches the Heisenberg QCNN$^+$ baseline of 0.839 at any shot count tested.}
            \label{fig:shot-sweep}
        \end{figure*}
        
    \subsection{Role of physics-informed design}\label{sec:inductive-bias}
        Incorporating Hamiltonian structure consistently improves performance within the QCNN family, but the same trend is not observed for shadow models, where the Hamiltonian-aware pre-rotation underperforms the strongest shadow baseline on every dataset. Among QCNNs, the Heisenberg variants outperform the standard architectures on every dataset. The improvement from QCNN Simple to Heisenberg QCNN is largest on $D_1$ ($0.647 \to 0.832$) and remains consistent on $D_3$ ($0.719 \to 0.810$), demonstrating that encoding the exchange structure of $H_0$ directly into the circuit provides a meaningful inductive bias~\cite{wiersema2020exploring, nguyen2024theory, larocca2022group}. The IsingXX, IsingYY, and IsingZZ gates generate the same interaction algebra as the Heisenberg coupling, encouraging the circuit to operate within physically relevant sectors of the low-energy subspace. The additional improvement from Heisenberg QCNN to Heisenberg QCNN+ is small ($\Delta R^2 \leq 0.015$ across $D_1$--$D_3$), indicating that additional expressibility beyond the exchange structure provides limited benefit for this task.
        
        The Hamiltonian Shadow CNN is consistently the weakest shadow variant across all five datasets, ranging from $R^2 = 0.241$ on $D_3$ to $0.833$ on $D_2$ at default shot budgets. The pre-rotation $e^{-i\tau H_0}$ is a global entangling unitary whose information about the eigenstructure of $H_0$ is distributed across multi-qubit correlations in the rotated state. The downstream CNN, however, extracts only single-site Pauli expectations from that rotated state, which may not capture information distributed into multi-qubit correlations by the global rotation. The fixed choice $\tau = 1.0$ may further limit performance by applying the same transformation regardless of quench strength or evolution time.
        
        This contrast points to a key difference between our two paradigms. In QCNNs, physics-informed structure enters through trainable gates that learn which sectors of the exchange algebra are relevant for each quench strength and evolution time. In shadow models, it enters as a fixed preprocessing step applied identically to all samples. These results suggest that physics-informed design is most effective when combined with trainable flexibility rather than applied as a fixed preprocessing step.

    \subsection{Effect of measurement budget}\label{sec:shot-scaling}
        Shadow performance generally improves with shot count across the datasets and protocols tested. Fig.~\ref{fig:shot-sweep} shows the sweep from 500 to 5{,}000~shots on $D_1$, $D_2$, and $D_3$.
        On $D_2$, all shadow models already exceed every QCNN at 500~shots and continue to improve, with Pauli Shadow reaching $R^2 = 0.934$ at 5{,}000~shots. On $D_3$, Clifford Shadow is the only variant to reach QCNN-level performance, achieving 0.847 at 5{,}000~shots, while Pauli and Hamiltonian shadows remain below all QCNNs at every shot level upto 5{,}000~shots, reaching only 0.612 and 0.489 at 5{,}000~shots. On $D_1$, no shadow model reaches the Heisenberg QCNN$^+$ baseline at any shot count tested. In the weak-quench regime, the advantage of physics-informed coherent circuits does not diminish with additional measurements. Per-$K$ breakdowns and a full resource comparison are given in Supplementary Fig.~\ref{fig:shot-sweep-grid} and Tables~\ref{tab:resources-training}--\ref{tab:resources-inference}.
        \begin{figure*}[t!]
            \centering
            \includegraphics[width=0.99\textwidth]{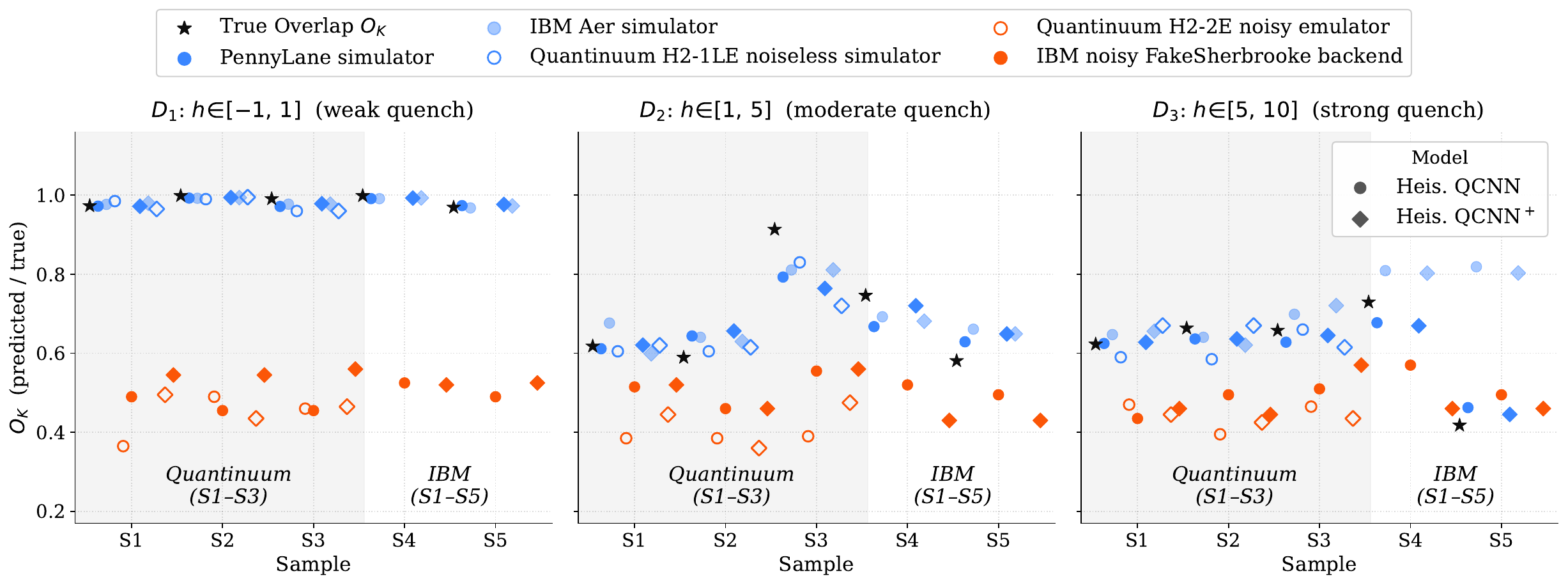}
            \caption{Per-sample noisy-backend predictions for Heisenberg QCNN (circles) and Heisenberg QCNN$^+$ (diamonds) at $K = 5$ across $D_1$, $D_2$, and $D_3$. The shaded region marks the three Quantinuum samples (S1 to S3) and IBM samples span all five (S1 to S5). Black stars show the true overlap $O_K$.  Blue marks noiseless predictions: filled opaque for PennyLane simulator, faded for IBM Aer simulator, and hollow for Quantinuum H2-1LE noiseless simulator. Orange marks noisy-backend predictions: hollow for Quantinuum H2-2E noisy emulator and filled opaque for IBM noisy FakeSherbrooke backend. Noiseless predictions from all three simulators closely track the ground truth, confirming correct circuit implementation. Predictions from both noisy backends cluster near 0.43 to 0.49 regardless of the true $O_K$, consistent with near-complete depolarization driven by the state preparation circuit ($2{,}044$ two-qubit gates).}
            \label{fig:hardware-per-sample}
        \end{figure*}
\section{Hardware Validation}\label{sec:hardware}
    We validate QCNN inference using two noisy backend models: a Quantinuum trapped-ion emulator and an IBM superconducting noise model. Since full-circuit inference is costly in terms of circuit execution and, in realistic hardware settings, job queue latency, we evaluate a representative subset at $K = 5$: three test samples from each of $D_1$, $D_2$, and $D_3$ on Quantinuum, and five test samples from each dataset on IBM, for the two physics-informed models (Heisenberg QCNN and Heisenberg QCNN+). Both studies run the full circuit, consisting of state preparation followed by the QCNN ansatz. All circuits are constructed using pytket~\cite{sivarajah2020tket} for Quantinuum and Qiskit~\cite{qiskit2024} for IBM. State preparation decomposes an arbitrary 10-qubit statevector into $2{,}044$ two-qubit gates regardless of entanglement structure. The QCNN ansatz requires an additional 70 to 150 two-qubit gates depending on the architecture.
    
    \subsection{Inference on Quantinuum trapped-ion backend}\label{sec:quantinuum}
        We first check circuit implementation on the H2-1LE noiseless simulator~\cite{moses2023race} (H1 gate set, all-to-all connectivity) using the two physics-informed QCNN architectures. Agreement with PennyLane is close: mean $|\Delta\langle Z\rangle| = 0.016$ on D1, $0.055$ on D2, and $0.072$ on D3, with residual differences consistent with shot noise at 200~shots. 
        
        We then run the full circuit on the H2-2E noisy emulator~\cite{moses2023race}, which models the Quantinuum H2-2 trapped-ion processor with realistic gate noise (${\sim}0.2\%$ two-qubit gate error rate). As shown in Fig.~\ref{fig:hardware-per-sample}, all H2-2E noisy-emulator predictions cluster near $0.43 \pm 0.04$ regardless of the true $O_K$ value, while PennyLane simulation predictions closely track the ground truth. The clustering near 0.43 is consistent with near-complete depolarization toward the maximally mixed state, which gives $(1 - \langle Z\rangle)/2 = 0.5$. The Heisenberg QCNN ansatz has an expected fidelity of approximately $(0.998)^{75} \approx 0.86$ on Quantinuum, where IsingXX/YY/ZZ gates are native. However, state preparation dominates the error budget: $(0.998)^{2044} \approx 0.017$.

    \subsection{Inference on IBM superconducting backend}\label{sec:ibm}
       We run the corresponding full-circuit experiment on the IBM FakeSherbrooke 127-qubit noise model via Qiskit Aer~\cite{qiskit2024}, which simulates the IBM Eagle superconducting processor with realistic gate noise (${\sim}1\%$ two-qubit error rate) and heavy-hex coupling constraints. An Aer statevector simulation without noise confirms correct implementation, with per-sample errors below 0.07 on D1 and D2.
        
        Under realistic IBM noise, FakeSherbrooke predictions also cluster near $0.49 \pm 0.03$, again consistent with near-complete depolarization (Fig.~\ref{fig:hardware-per-sample}). The collapse is driven by the same bottleneck as on Quantinuum: state preparation requires ${\sim}2{,}044$ two-qubit gates, yielding $(0.99)^{2044} \approx 0$ expected fidelity at IBM's 1\% two-qubit error rate. The QCNN ansatz contributes an additional factor of approximately only $(0.99)^{150} \approx 0.22$.

    \subsection{Hardware Inference Analysis}\label{sec:bottleneck}
        Both noisy-backend studies identify arbitrary state preparation needed for training as the dominant noise source. This bottleneck is not specific to QCNNs: any hardware implementation that begins from a classically specified statevector $\ket{\psi(t)}$ would require preparing that state before either QCNN inference or shadow measurement. In practical deployment, $\ket{\psi(t)}$ would instead be generated directly on the device through physical time evolution under $H$. Trotterized time evolution~\cite{lloyd1996universal} or a variational approach for the 10-qubit Heisenberg chain will require substantially fewer two-qubit gates than arbitrary state preparation, improving the expected preparation fidelity on both backends. This suggests a more favorable path on Quantinuum with physically motivated state generation, while IBM would additionally require improvements in gate fidelity or error mitigation techniques~\cite{temme2017error}.

\section{Discussion and conclusion}\label{sec:conclusion}
    We have investigated two quantum information extraction strategies for predicting low-energy subspace overlaps in a Heisenberg spin chain following a local quench: measurement-based classical shadow tomography and coherent quantum convolutional neural networks. Both families successfully learn the target property across five dataset configurations, supporting quantum information extraction as a viable paradigm for learning many-body properties. The optimal strategy depends on the quench regime rather than a universal preference for one approach over the other. Physics-informed QCNNs achieve stable performance across all five datasets, while shadow methods perform best under moderate perturbations. Incorporating the Heisenberg exchange into QCNN gates provides consistent gains, while the same prior applied as a fixed pre-rotation in the shadow protocol does not.
    
   Noisy-backend experiments using Quantinuum and IBM models show that arbitrary state preparation is the dominant noise source. Loading a classically specified statevector requires thousands of two-qubit gates, and current error rates cause near-complete depolarization before inference begins. In practical deployment, however, physical time evolution under $H$ would generate the quantum state directly on the device, eliminating the need for arbitrary state preparation. Trotterized circuits and variational state preparation are promising alternatives when the time-evolved state must be prepared.
    
   The present study focuses on a 10-qubit system with a specific Hamiltonian structure, and scaling to larger systems is a natural next step. The learning strategies studied in this work are applicable to other many-body systems where subspace overlaps are physically relevant, such as transverse-field Ising chains, Fermi-Hubbard models, and frustrated spin systems. Future work should also validate these methods on real quantum hardware using physically generated time-evolved states. Whether physics-informed designs, including Hamiltonian-aware shadow protocols and exchange-aligned QCNN gates tailored to the target Hamiltonian, provide advantages in these settings remains an important open question.

\section{Acknowledgements}
    The research presented in this article was supported by the NNSA’s Advanced Simulation and Computing Beyond Moore’s Law program at Los Alamos National Laboratory and the Laboratory Directed Research and Development program of Los Alamos National Laboratory under project number 20260043DR. 
    This work has been assigned LANL technical report number LA-UR-26-25363.

\section*{Supplementary Information}
    \subsection*{Heisenberg QCNN Gate Design}
    \label{supp:ising-gates}
    
    The convolutional gates of the Heisenberg QCNN directly reflect the nearest-neighbor exchange structure of $H_0$. Each exchange term in Eq.~\eqref{eq:H0} generates a two-qubit Ising interaction whose unitary is
    \begin{align}
        \label{eq:ising_unitary}
        e^{-i\theta P_i P_{i+1}/2} &= 
        \cos\tfrac{\theta}{2}\, I \otimes I 
        - i\sin\tfrac{\theta}{2}\, P_i \otimes P_{i+1},\nonumber\\
        \span\quad P \in \{X,Y,Z\}.
    \end{align}
    In the two-qubit computational basis, the corresponding $XX$ rotation takes the form
    \begin{equation}
        \label{eq:ising_matrix}
        \mathrm{IsingXX}(\theta) = 
        \begin{pmatrix}
        \cos\frac{\theta}{2} & 0 & 0 & -i\sin\frac{\theta}{2} \\
        0 & \cos\frac{\theta}{2} & -i\sin\frac{\theta}{2} & 0 \\
        0 & -i\sin\frac{\theta}{2} & \cos\frac{\theta}{2} & 0 \\
        -i\sin\frac{\theta}{2} & 0 & 0 & \cos\frac{\theta}{2}
        \end{pmatrix},
    \end{equation}
    with analogous forms for $\mathrm{IsingYY}(\beta)$ and $\mathrm{IsingZZ}(\gamma)$. The three gates together span all three exchange channels of the Heisenberg interaction in Eq.~\eqref{eq:H0}, generating the same operator algebra as the nearest-neighbor coupling terms of $H_0$. Each gate parameter $(\alpha, \beta, \gamma)$ is independently trainable, allowing the circuit to learn the relative weights of the three exchange channels appropriate for each quench regime. This provides the inductive bias of the Heisenberg QCNN: rather than using generic two-qubit blocks, optimization is restricted to a physically motivated subspace spanned by the exchange interactions of $H_0$.

    \subsection*{Training Details}
    \label{supp:training}
        \paragraph{Hyperparameters.}
        Table~\ref{tab:resources-hyperparams} summarizes all hyperparameters used for both model families. All models train for a maximum of 500 epochs with a ReduceLROnPlateau scheduler (factor 0.5, patience 15 epochs, minimum learning rate $10^{-4}$) and early stopping with patience 30 epochs on validation MSE. The model with the lowest validation MSE is restored before evaluation.

        \begin{table}[t]
            \centering
            \caption{Training hyperparameters for shadow CNNs and QCNNs.}
            \label{tab:resources-hyperparams}
            \begin{tabular}{@{}lll@{}}
                \toprule
                Hyperparameter & Shadow CNN & QCNN \\
                \midrule
                Optimizer            & Adam & Adam \\
                Learning rate        & $10^{-3}$ & $10^{-2}$ \\
                Weight decay         & $10^{-4}$ & --- \\
                Loss function        & MSE & Huber ($\delta = 0.05$) \\
                Batch size           & 32 & 16 \\
                Max epochs           & 500 & 500 \\
                Early stop patience  & 30 & 30 \\
                Scheduler factor     & 0.5 & 0.5 \\
                Scheduler patience   & 15 & 15 \\
                Min learning rate    & $10^{-4}$ & $10^{-4}$ \\
                Parameter init       & PyTorch default & $\mathcal{N}(0,\,0.01^2)$ \\
                \bottomrule
            \end{tabular}
        \end{table}

        \paragraph{Target transformation.}
        All models predict the residual $1 - O_K$ during training and transform back for evaluation. This concentrates targets near zero in the weak-quench regime where $O_K \approx 1$, reducing the effective dynamic range and improving gradient flow. Both model families produce outputs naturally bounded in $[0, 1]$: QCNNs via the $(1 - \langle Z \rangle)/2$ mapping, and shadow CNNs via a sigmoid output activation.

        \paragraph{Gradient computation.}
        QCNN gradients are computed via adjoint differentiation~\cite{jones2020efficient}, which requires only two forward passes through the circuit regardless of parameter count, compared to the $2p$ evaluations required by the parameter-shift rule~\cite{mitarai2018quantum, schuld2019evaluating}. For circuits with $p > 100$ parameters, this substantially reduces the number of circuit evaluations required for training. The \texttt{lightning.qubit} simulator~\cite{bergholm2018pennylane} is used when available, falling back to \texttt{default.qubit} otherwise.

        \paragraph{Evaluation metric.}
        Performance is measured by the coefficient of determination ($R^2$) on the held-out test set:
            \begin{equation}\label{eq:r2-supp}
              R^2 = 1 - \frac{\sum_i (y_i - \hat{y}_i)^2}
                              {\sum_i (y_i - \bar{y})^2}\,,
            \end{equation}
        where $y_i$ are the true overlap values, $\hat{y}_i$ the model predictions, and $\bar{y}$ the test-set mean. An $R^2$ of 1 indicates perfect prediction; $R^2 = 0$ corresponds to predicting the mean for all inputs. We report mean $R^2$ across three independent runs, with error bars indicating $\pm 1$ standard deviation. The dataset-level summary $R^2$ denotes the mean across all $K$ values in that dataset's $K$-set.
    \begin{table}[t!]
        \centering
        \caption{Mean $R^2$ averaged over all $K$ values (default shots:
        500 for Pauli and Hamiltonian shadows, 1{,}000 for Clifford shadow).
        Results are shown as means across three random seeds.}
        \label{tab:sup-results-summary}
        \begin{tabular}{@{}lccccc@{}}
            \toprule
            Model & D1 & D2 & D3 & D4 & D5 \\
            \midrule
            \multicolumn{6}{l}{\textit{Shadow + Classical CNNs}} \\[2pt]
            Pauli Shadow CNN       & 0.615 & 0.886 & 0.332 & 0.567 & 0.599 \\
            Clifford Shadow CNN    & 0.508 & 0.861 & 0.672 & 0.704 & 0.792 \\
            Hamiltonian Shadow CNN & 0.495 & 0.833 & 0.241 & 0.497 & 0.403 \\
            \midrule
            \multicolumn{6}{l}{\textit{QCNNs}} \\[2pt]
            QCNN Simple            & 0.647 & 0.632 & 0.719 & 0.743 & 0.715 \\
            QCNN Deep              & 0.713 & 0.713 & 0.777 & 0.770 & 0.809 \\
            Heisenberg QCNN        & 0.832 & 0.753 & 0.810 & 0.828 & 0.830 \\
            Heisenberg QCNN+       & 0.839 & 0.767 & 0.817 & 0.840 & 0.846 \\
            \bottomrule
        \end{tabular}
    \end{table}

    \begin{table}[t!]
        \centering
        \caption{Shadow model $R^2$ as a function of shot count on
        $D_1$--$D_3$, with Heisenberg QCNN baselines for comparison.}
        \label{tab:shot-sweep}
        \begin{tabular}{@{}llcccc@{}}
            \toprule
            Dataset & Model & 500 & 1000 & 2000 & 5000 \\
            \midrule
            D1 & Pauli       & 0.615 & 0.601 & 0.659 & 0.773 \\
               & Clifford    & 0.464 & 0.508 & 0.543 & 0.665 \\
               & Hamiltonian & 0.495 & 0.587 & 0.688 & 0.727 \\
            \cmidrule{2-6}
               & \multicolumn{5}{l}{\small Heis.\ QCNN 0.832\quad
                 Heis.\ QCNN+ 0.839} \\
            \midrule
            D2 & Pauli       & 0.886 & 0.900 & 0.925 & 0.934 \\
               & Clifford    & 0.815 & 0.861 & 0.893 & 0.918 \\
               & Hamiltonian & 0.833 & 0.856 & 0.871 & 0.893 \\
            \cmidrule{2-6}
               & \multicolumn{5}{l}{\small Heis.\ QCNN 0.753\quad
                 Heis.\ QCNN+ 0.767} \\
            \midrule
            D3 & Pauli       & 0.332 & 0.449 & 0.517 & 0.612 \\
               & Clifford    & 0.475 & 0.672 & 0.749 & 0.847 \\
               & Hamiltonian & 0.241 & 0.349 & 0.415 & 0.489 \\
            \cmidrule{2-6}
               & \multicolumn{5}{l}{\small Heis.\ QCNN 0.810\quad
                 Heis.\ QCNN+ 0.817} \\
            \bottomrule
        \end{tabular}
    \end{table}
    \paragraph{Parallelization.}
        Training is parallelized across $K$ values using MPI via \texttt{mpi4py}~\cite{dalcin2011parallel}. Each MPI rank trains an independent subset of the $K$-set with no inter-rank communication during training. Shadow features are precomputed once on all ranks before training begins and shared across all $K$ values, avoiding redundant computation. Results are gathered on rank 0 after all ranks complete.

    \begin{table*}[t!]
        \centering
        \caption{Quantum resource cost for training. Shadow costs are the one-time feature extraction over $S_{\mathrm{train}} = 240$ samples; CNN optimization is purely classical. QCNN costs assume ${\sim}350$ effective epochs (early stopping from 500 max) with adjoint differentiation (${\approx}3$ circuit executions per gradient step per sample); validation monitoring is classical and excluded.
        All gates decomposed into 1Q rotations and CNOTs ($\mathrm{IsingXX/ZZ} \to 1\text{ 1Q} + 2\text{ CNOT}$; $\mathrm{IsingYY} \to 5\text{ 1Q} + 2\text{ CNOT}$). State preparation of $\ket{\psi(t)}$ is excluded; on hardware it would be realized by physical time evolution under~$H$.}
        \label{tab:resources-training}
       
        \begin{tabular}{@{}lrrrrrr@{}}
            \toprule
            \textbf{Model} & \textbf{Shots} & \textbf{Params} &
            \textbf{Circuit Exec.} & \textbf{1Q Gates} & \textbf{CNOTs} &
            \textbf{Meas.} \\
            \midrule
            \multicolumn{7}{l}{\textit{Shadow + Classical CNNs}} \\[2pt]
            Pauli Shadow CNN       &      500 & 42\,241 &     120\,000 &     1\,200\,000 &              0 &   1\,200\,000 \\
            Pauli Shadow CNN       &    1\,000 & 42\,241 &     240\,000 &     2\,400\,000 &              0 &   2\,400\,000 \\
            Pauli Shadow CNN       &    2\,000 & 42\,241 &     480\,000 &     4\,800\,000 &              0 &   4\,800\,000 \\
            Pauli Shadow CNN       &    5\,000 & 42\,241 &   1\,200\,000 &    12\,000\,000 &              0 &  12\,000\,000 \\[3pt]
            Clifford Shadow CNN    &      500 & 42\,529 &     120\,000 &     9\,720\,000 &    5\,670\,000 &   1\,200\,000 \\
            Clifford Shadow CNN    &    1\,000 & 42\,529 &     240\,000 &    19\,440\,000 &   11\,340\,000 &   2\,400\,000 \\
            Clifford Shadow CNN    &    2\,000 & 42\,529 &     480\,000 &    38\,880\,000 &   22\,680\,000 &   4\,800\,000 \\
            Clifford Shadow CNN    &    5\,000 & 42\,529 &   1\,200\,000 &    97\,200\,000 &   56\,700\,000 &  12\,000\,000 \\[3pt]
            Hamiltonian Shadow CNN &      500 & 42\,241 &     120\,000 &     1\,217\,520 &       12\,960 &   1\,200\,000 \\
            Hamiltonian Shadow CNN &    1\,000 & 42\,241 &     240\,000 &     2\,417\,520 &       12\,960 &   2\,400\,000 \\
            Hamiltonian Shadow CNN &    2\,000 & 42\,241 &     480\,000 &     4\,817\,520 &       12\,960 &   4\,800\,000 \\
            Hamiltonian Shadow CNN &    5\,000 & 42\,241 &   1\,200\,000 &    12\,017\,520 &       12\,960 &  12\,000\,000 \\
            \midrule
            \multicolumn{7}{l}{\textit{QCNNs}} \\[2pt]
            QCNN Simple            &        1 &     123 &     252\,000 &    42\,336\,000 &   17\,640\,000 &     252\,000 \\
            QCNN Deep              &        1 &     219 &     252\,000 &    66\,528\,000 &   21\,672\,000 &     252\,000 \\
            Heisenberg QCNN        &        1 &     148 &     252\,000 &    62\,496\,000 &   37\,800\,000 &     252\,000 \\
            Heisenberg QCNN+       &        1 &     212 &     252\,000 &    78\,624\,000 &   37\,800\,000 &     252\,000 \\
            \bottomrule
        \end{tabular}
    \end{table*}
    \subsection*{Resource Profiles}\label{sec:resources}
    
   Supplementary Table~\ref{tab:resources-training} reveals qualitatively different resource profiles for model training. Shadow models pay their quantum cost once during feature extraction: at 500~shots, Pauli Shadow CNN requires only 1.2M single-qubit gates and zero CNOTs to process the 240~training samples, after which CNN optimization is entirely classical. In contrast, even the simplest QCNN requires ${\sim}252{,}000$ circuit executions totaling 42.3M single-qubit gates and 17.6M CNOTs for ${\sim}350$ epochs of adjoint-differentiation training. At default shots, shadow training requires $35$--$66\times$ fewer single-qubit gates than QCNN training.

    \begin{table*}[t!]
        \centering
        \caption{Quantum resource cost per prediction (inference). Shadow models require $S$ shots of randomized measurements to extract features, then a single classical CNN forward pass. QCNNs require one circuit execution and one measurement. Gate decomposition and state-preparation conventions follow Table~\ref{tab:resources-training}.}
        \label{tab:resources-inference}
        
        \begin{tabular}{@{}lrrrrrr@{}}
            \toprule
            \textbf{Model} & \textbf{Shots} & \textbf{Params} &
            \textbf{Circuit Exec.} & \textbf{1Q Gates} & \textbf{CNOTs} &
            \textbf{Meas.} \\
            \midrule
            \multicolumn{7}{l}{\textit{Shadow + Classical CNNs}} \\[2pt]
            Pauli Shadow CNN       &      500 & 42\,241 &        500 &      5\,000 &          0 &    5\,000 \\
            Pauli Shadow CNN       &    1\,000 & 42\,241 &      1\,000 &     10\,000 &          0 &   10\,000 \\
            Pauli Shadow CNN       &    2\,000 & 42\,241 &      2\,000 &     20\,000 &          0 &   20\,000 \\
            Pauli Shadow CNN       &    5\,000 & 42\,241 &      5\,000 &     50\,000 &          0 &   50\,000 \\[3pt]
            Clifford Shadow CNN    &      500 & 42\,529 &        500 &     40\,500 &     23\,625 &    5\,000 \\
            Clifford Shadow CNN    &    1\,000 & 42\,529 &      1\,000 &     81\,000 &     47\,250 &   10\,000 \\
            Clifford Shadow CNN    &    2\,000 & 42\,529 &      2\,000 &    162\,000 &     94\,500 &   20\,000 \\
            Clifford Shadow CNN    &    5\,000 & 42\,529 &      5\,000 &    405\,000 &    236\,250 &   50\,000 \\[3pt]
            Hamiltonian Shadow CNN &      500 & 42\,241 &        500 &      5\,073 &         54 &    5\,000 \\
            Hamiltonian Shadow CNN &    1\,000 & 42\,241 &      1\,000 &     10\,073 &         54 &   10\,000 \\
            Hamiltonian Shadow CNN &    2\,000 & 42\,241 &      2\,000 &     20\,073 &         54 &   20\,000 \\
            Hamiltonian Shadow CNN &    5\,000 & 42\,241 &      5\,000 &     50\,073 &         54 &   50\,000 \\
            \midrule
            \multicolumn{7}{l}{\textit{QCNNs}} \\[2pt]
            QCNN Simple            &        1 &     123 &          1 &        168 &         70 &        1 \\
            QCNN Deep              &        1 &     219 &          1 &        264 &         86 &        1 \\
            Heisenberg QCNN        &        1 &     148 &          1 &        248 &        150 &        1 \\
            Heisenberg QCNN+       &        1 &     212 &          1 &        312 &        150 &        1 \\
            \bottomrule
        \end{tabular}
    \end{table*}
        
    For inference, Supplementary Table~\ref{tab:resources-inference} shows that the resource profile reverses. At default shots, Clifford Shadow CNN requires 47{,}250~CNOTs per prediction, which is $315\times$ the Heisenberg QCNN's 150~CNOTs, along with 81{,}000 single-qubit gates (vs.\ 248) and 42{,}529 classical parameters (vs.\ 148). QCNNs thus provide a compact representation: 148~parameters encode predictive structure that shadow methods distribute across many measurement circuits, consistent with the expectation that parametrized circuits can learn compressed representations of quantum properties~\cite{cong2019quantum, pesah2021absence}.
    
    Whether these different profiles are favorable in practice depends on the hardware constraints: shadow measurements are fully parallel and tolerant of gate errors~\cite{chen2021robust}, whereas QCNNs require coherent execution of the full circuit depth. In regimes where shadow models match or exceed QCNN accuracy (D2 at any shot count; D3 Clifford at 5{,}000~shots), the cost is paid on the measurement side via many independent shallow circuits. In regimes where QCNNs maintain their advantage (D1 weak quench), the physics-informed QCNN achieves superior accuracy with orders of magnitude fewer quantum resources per prediction.

        \subsection*{Shot Sweep: Per-\texorpdfstring{$K$}{K} Breakdown}
    \label{supp:shot-sweep}
    
\begin{figure*}[t!]
            \centering
            \includegraphics[width=0.99\textwidth]{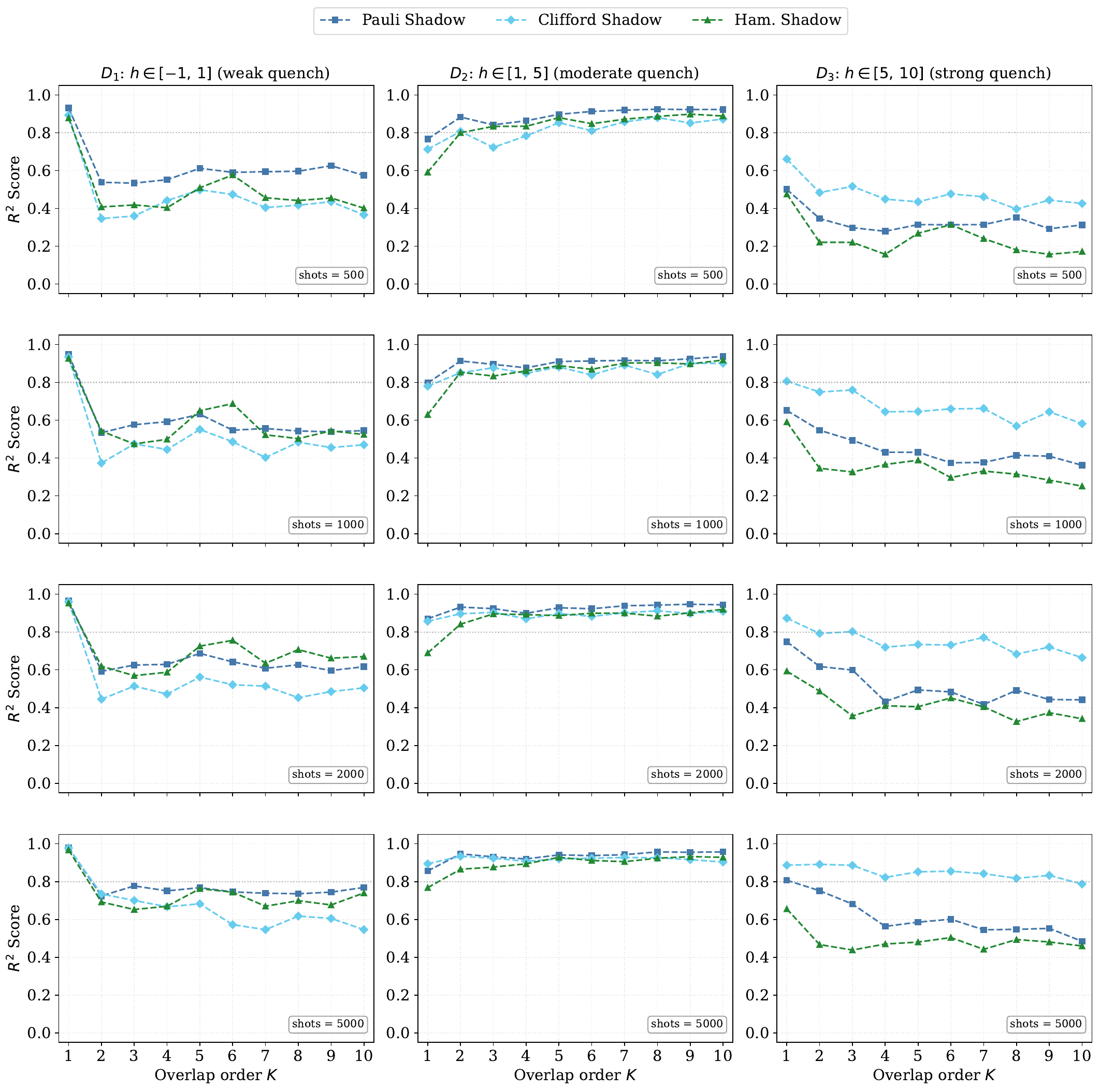}
            \caption{Test-set $R^2$ as a function of overlap order $K$ for the three shadow models at each shot level (500, 1000, 2000, 5000 shots), shown across $D_1$ (weak quench), $D_2$ (moderate quench), and $D_3$ (strong quench). Each row corresponds to one shot level and each column to one dataset. The gray dotted line marks $R^2 = 0.8$. On $D_2$, all three shadow models exceed 0.8 at 500 shots and improve consistently with increasing shot count. On $D_3$, Clifford Shadow crosses 0.8 only at 2000 shots and above, while Pauli and Hamiltonian shadows remain below 0.8 at all shot levels tested. On $D_1$, no shadow model exceeds 0.8 at any shot count.}
            \label{fig:shot-sweep-grid}
    \end{figure*}

    Fig.~\ref{fig:shot-sweep-grid} shows test-set $R^2$ as a function of overlap order $K$ for all three shadow models at different shot levels across $D_1$, $D_2$, and $D_3$. On $D_2$, all three shadow models exceed $R^2 = 0.8$ at 500 shots across nearly all $K$ values and improve consistently with increasing shot count. Performance is relatively uniform across $K$, indicating that the moderate-quench regime is well-captured by shadow features regardless of the overlap order. On $D_3$, the per-$K$ curves reveal substantial variation. Clifford Shadow crosses the 0.8 threshold only at 2000 shots and above, primarily for smaller $K$ values where the subspace overlap appears easier to learn from the measured correlators. Pauli and Hamiltonian shadows remain below 0.8 at all shot levels and all $K$ values tested, with performance degrading further at larger $K$ where the variance of $O_K$ shrinks and prediction errors are amplified in the $R^2$ denominator. On $D_1$, no shadow model exceeds 0.8 at any shot count or $K$ value. The gap between shadow models and Heisenberg QCNN+ (0.839) is consistent across $K$, suggesting that the limitation is not purely statistical: local shadow features may not capture the near-perturbative structure that physics-informed coherent circuits exploit through their Hamiltonian-aligned gates.

\bibliographystyle{apsrev4-2} 
\showtitleinbib
\bibliography{fixapsbib,main}
\end{document}